 \documentclass[10pt]{emulateapj}

\slugcomment{Accepted for publication in The Astrophysical 
             Journal (Date: July 9, 2007)}

\shorttitle{UNSTABLE DISK GALAXIES. I. MODAL PROPERTIES}
\shortauthors{M. A. Jalali}

\begin{document}

\title{UNSTABLE DISK GALAXIES. I. MODAL PROPERTIES}

\author{Mir Abbas Jalali
}
\affil{Sharif University of Technology, Azadi Avenue, Tehran, Iran;
mjalali@sharif.edu}


\begin{abstract}

I utilize the Petrov-Galerkin formulation and develop a new method
for solving the unsteady collisionless Boltzmann equation in both
the linear and nonlinear regimes. In the first order approximation, 
the method reduces to a linear eigenvalue problem which is solved
using standard numerical methods. I apply the method to the dynamics
of a model stellar disk which is embedded in the field of a
soft-centered logarithmic potential. The outcome is the full
spectrum of eigenfrequencies and their conjugate normal modes for
prescribed azimuthal wavenumbers. The results show that the
fundamental bar mode is isolated in the frequency space while
spiral modes belong to discrete families that bifurcate from the
continuous family of van Kampen modes. The population of spiral
modes in the bifurcating family increases by cooling the disk and
declines by increasing the fraction of dark to luminous matter. It
is shown that the variety of unstable modes is controlled by the
shape of the dark matter density profile. 
\end{abstract} 


\section{INTRODUCTION}

Dynamics of self-gravitating stellar systems and plasma fluids
are governed by the collisionless Boltzmann equation (CBE) \citep{BT87}.
Finding a general solution of the CBE has been a challenging problem
in various disciplines of physical sciences. Due to existing difficulties
of the general problem, finding a solution to the linearized CBE became
the center of attraction in the twentieth century when \citet{L46} and
van Kampen (1955) discovered the normal modes of collisionless ensembles.
Later in 1970's, \citet{K71,K77} developed a matrix theory that was
capable of computing normal modes of stellar systems through solving a
nonlinear eigenvalue problem. His theory remained as the only analytical
perturbation theory used by the community of galactic dynamicists over
the past three decades.

\citet{K77} assumed an exponential form
$\exp(-{\rm i}\omega t)$ for the time-dependent part
of physical quantities where ${\rm i}=\sqrt{-1}$, and
solved the linearized CBE for the perturbed distribution
function (DF) $f_1$ in terms of the perturbed
potential $V_1$. After expanding the potential and
density functions in terms of bi-orthogonal basis sets
in the configuration space, he used the weighted residual
form of the fundamental equation
\begin{equation}
f_1 d \textbf{\textit{v}} d \textbf{\textit{x}}=
\Sigma_1 d \textbf{\textit{x}},
\end{equation}
to obtain a nonlinear eigenvalue problem for the complex
eigenfrequency $\omega$. For self-consistent perturbations
the surface density $\Sigma_1=\int f_1 d \textbf{\textit{v}}$ is
related to $V_1$ through Poisson's integral, and the symbols
$ d \textbf{\textit{v}}$ and $ d \textbf{\textit{x}}$ denote the
elements of velocity and position vectors.

\citet{Z76} used Kalnajs's theory to compute the modes of the
isothermal disk \citep{M63}, which has the astrophysically important
property of a flat rotation curve. His analysis was then extended by
\citet{ER98a,ER98b} to general scale-free disks with arbitrary cusp
slopes. Application of Kalnajs's theory to soft-centered models of
stellar disks has been mainly focused on the isochrone and
Kuzmin-Toomre disks \citep{K78,H92,PC97}.
A disk with exponential light profile and an approximately
flat rotation curve was also investigated by \citet{VD96}.
More recently Jalali \& Hunter (2005a, hereafter JH) gave
new results for soft-centered models of stellar disks. They showed
the importance of a boundary integral in the modal properties of
unidirectional disks and computed a fundamental bar mode and a
secondary spiral mode for the isochrone, Kuzmin-Toomre and a newly
introduced family of cored exponential disks. JH also extended Kalnajs's
first order perturbation theory to the second order, and illustrated
energy and angular momentum content of different Fourier components.
Their bar charts showed that only a few number of expansion terms in
the radial angle govern the perturbed dynamics.

Implementation of Kalnajs's (1977) theory, however, has some technical
problems due to the nonlinear dependency of his matrix equations
on $\omega$. Most computational methods that deal with nonlinear
eigenvalue problems are iterative. They start with an initial guess
of the solution and continue with a search scheme in the frequency
space. Newton's method is perhaps the most efficient technique that
guarantees a quadratic convergence should the initial guess be close
enough to the solution. The key issue in success of any iterative
scheme is the attracting or repelling nature of an eigenvalue.
It is obvious that only attracting eigenvalues can be captured by
iterative methods while we have no priori knowledge of their
basins of attraction in order to make our initial guess.
The mentioned computational difficulties make it a formidable task 
to explore all normal modes of a stellar system, which include 
growing modes as well as stationary van Kampen modes. Moreover, 
it is not easy to develop a general nonlinear theory based on 
Kalnajs's method for studying the interaction of modes. 

\citet{P04,P05} introduced an alternative method for the normal mode
calculation of stellar disks whose outcome was a linear eigenvalue
problem for $\omega$. His method is capable of finding all
eigenmodes of a stellar disk should one use fine grids in the action
space. Polyachenko's method is somehow costly because it results in
a large linear system of equations to assure point-wise convergence
in the action space and mean convergence of Fourier expansions in
the space of the radial angle. Extension of his method to nonlinear
regime is another challenging problem yet to be investigated.
\citet{T05} has also followed an approach similar to \citet{P05} and
studied the instability of stellar disks surrounding massive
objects. His eigenvalue equations involve action variables, and
practically, need to be solved over a discretized grid in the space
of actions.

Recent developments in fluid mechanics \citep{DG95,MS99} inspired me
to formulate the dynamics of stellar systems in a new framework,
which is capable of solving the CBE not only in the linear regime,
but also in its full nonlinear form. The method systematically
searches for smooth solutions of the CBE by expanding the perturbed
DF using Fourier series of angle variables and an appropriate set of
trial functions in the space of actions. Coefficients of expansion
are unknown time-dependent amplitude functions whose evolution
equations are obtained by the Petrov-Galerkin projection \citep{F72} 
of the CBE. That is indeed taking the weighted residual form of the 
CBE by integration over the action-angle space and deriving a system 
of nonlinear ordinary differential equations (ODEs) for the amplitude
functions. The associated first order system of ODEs leads to a
linear eigenvalue problem, which is solved using standard numerical
methods.
 
In this paper I present my new method and apply it to explore the
modal properties of a model galaxy. In a second paper, I will
address the nonlinear evolution of modes and wave interactions. The
paper is organized as follows. In sections
\ref{sec::dynamical-theory} and \ref{sec::linear-theory}, I use the
Petrov-Galerkin method to project the CBE to a system of ODEs in the
time domain and derive a system of linear eigenvalue equations. Section
\ref{sec:modes-exp-disk} presents the eigenfrequency spectra and
their corresponding mode shapes of the cored exponential disk of JH.
The stars of this model move in the field of a soft-centered
logarithmic potential. I also investigate the effect of physical
parameters of the equilibrium model on the modal content. 
In section \ref{sec:final-discussions}, I discuss on the nature of 
a self-gravitating mode and compare the performance of my method 
with other theories. Some fundamental achievements of this work 
are summarized in section \ref{sec:conclusions}.

\section{NONLINEAR THEORY}
\label{sec::dynamical-theory}

I use the usual polar coordinates $\textbf{\textit{x}}=(R,\phi)$ and assume
that the temporal evolution of the DF and gravitational potential
starts from an axisymmetric equilibrium state described by
$f_0(\textbf{\textit{x}},\textbf{\textit{p}})$ and $V_0(R)$ so that
\begin{eqnarray}
f(\textbf{\textit{x}},\textbf{\textit{p}},t) &=&
f_0(\textbf{\textit{x}},\textbf{\textit{p}})+
f_1(\textbf{\textit{x}},\textbf{\textit{p}},t), \\
V(\textbf{\textit{x}},\textbf{\textit{p}},t) &=&
V_0(R)+V_1(\textbf{\textit{x}},\textbf{\textit{p}},t).
\label{eq:perturbations}
\end{eqnarray}
Here $\textbf{\textit{p}}=\left (p_R,p_\phi \right )$ is the
momentum vector conjugate to $\textbf{\textit{x}}=\left (R,\phi \right )$.
Motion of stars in the equilibrium state is governed by the
zeroth order Hamiltonian
\begin{equation}
{\cal H}_0=\frac 12 \left ( p_R^2+\frac {p_\phi^2}{R^2}\right
)+V_0(R). \label{eq:zeroth-order-H}
\end{equation}
For bounded orbits, $R$ and $\phi$ become librating and rotating,
respectively. One can therefore describe the dynamics using the
action variables $\textbf{\textit{J}}=\left (J_R,J_\phi \right )$,
\begin{equation}
J_R=\oint p_R  d R,~~J_\phi=\oint p_\phi
d \phi=p_\phi,\label{eq:define-actions}
\end{equation}
and their conjugate angles $\Theta=(\theta_R,\theta_\phi)$. A
transformation $(\textbf{\textit{x}},\textbf{\textit{p}})\rightarrow
(\Theta,\textbf{\textit{J}})$ leaves the Hamiltonian ${\cal H}_0$ as
a function of actions only, ${\cal H}_0(\textbf{\textit{J}})$, and
therefore, the phase space flows of the equilibrium state lie on a
two dimensional torus $\textbf{\textit{J}}=\textbf{\textit{c}}$ with
$\textbf{\textit{c}}$ being a constant 2-vector. An action-angle
transformation can locally be found for any bounded regular orbit,
but it is a global transformation if only one orbit family occupies
the phase space. The axisymmetric potential $V_0(R)$ supports only
rosette orbits. Radial and circular orbits are the limiting cases of
rosette orbits with $J_\phi=0$ and $J_R=0$, respectively. By
representing $f$ in terms of the action-angle variables, the CBE
reads
\begin{equation}
\frac{\partial f}{\partial t}+[ f,{\cal H} ]=0,
\label{eq:CBE-action-angle}
\end{equation}
where $[,]$ denotes a Poisson bracket taken over the action-angle
space. According to Jeans theorem \citep{J1915,L-B62} $f_0$ depends
on the phase space coordinates through the integrals of motion,
which are the actions in the present formulation, and one obtains
$[f_0,{\cal H}_0]=0$. Subsequently, equation (\ref{eq:CBE-action-angle})
may be rewritten as
\begin{equation}
\frac{\partial f_1}{\partial t}=-\left [f_1,{\cal H}_0 \right ]-
\left [ f_0,{\cal H}_1 \right ]-\left [f_1,{\cal H}_1 \right ],
\label{eq:CBE-perturbed}
\end{equation}
where ${\cal H}_1$ is the perturbed Hamiltonian. A dark matter halo
contributes both to ${\cal H}_0$ and to ${\cal H}_1$ if it is live,
i.e., if it exchanges momentum/energy with the luminous stellar component.
In this paper I confine myself to a rigid halo that only contributes
to ${\cal H}_0$ through $V_0$ and assume that ${\cal H}_1=V_1$ is the
perturbed potential due to self-gravity.

Let me expand $f_1$ and $V_1$ in Fourier series of angle variables
and write
\begin{eqnarray}
f_1(\Theta,\textbf{\textit{J}},t) &=& \!\!
\sum_{m,l=-\infty}^{\infty}\sum_{j=0}^{\infty} d^{ml}_j(t)
\Phi^{ml}_j(\textbf{\textit{J}})e^{{\rm i}\left(m\theta_\phi+l\theta_R
\right)},
\label{eq:expansion-f1} \\
V_1(\Theta,\textbf{\textit{J}},t) &=& \!\!
\sum_{m,l=-\infty}^{\infty}\sum_{j=0}^{\infty} b^{ml}_j(t)
\Psi^{ml}_j(\textbf{\textit{J}})e^{{\rm i}\left(m\theta_\phi+l\theta_R
\right)}, \label{eq:expansion-V1}
\end{eqnarray}
where $\Phi^{ml}_j(\textbf{\textit{J}})$ and
$\Psi^{ml}_j(\textbf{\textit{J}})$ are
some trial functions in the space of action variables, and
$d^{ml}_j(t)$ and $b^{ml}_j(t)$ are time-dependent amplitude
functions. On the other hand, one can expand $V_1$ and its
corresponding surface density $\Sigma_1$ in the configuration space
as
\begin{eqnarray}
\Sigma_1(R,\phi,t) &=& \!\!
\sum_{m=-\infty}^{\infty}\sum_{j=0}^{\infty} a^m_j(t)
\sigma^{|m|}_j(R)e^{{\rm i}m\phi},
\label{eq:expansion-sigma1-config} \\
V_1(R,\phi,t) &=& \!\! \sum_{m=-\infty}^{\infty}\sum_{j=0}^{\infty}
a^m_j(t) \psi^{|m|}_j(R)e^{{\rm i}m\phi}. \label{eq:expansion-V1-config}
\end{eqnarray}
Here $\psi^{|m|}_j(R)$ and $\sigma^{|m|}_j(R)$ are bi-orthogonal
potential--surface density pairs that satisfy the relation
\begin{equation}
\label{eq:Djk-definition}
2\pi \int\limits_{0}^{\infty}
\psi^{|m|}_j(R)\sigma^{|m'|}_{j'}(R)R d R = D_j(m)\delta
_{m,m'}\delta_{j,j'},
\end{equation}
where $\delta_{m,m'}$ is the Kronecker delta and $D_j(m)$ are some
constants. It is remarked that the real parts of $\Sigma_1$, $V_1$
and $f_1$ describe physical solutions. In this work I utilize the
\citet{CB72} functions
\begin{eqnarray}
\psi^{|m|}_j &=&
      -\frac {1}{b} \left (\! {1-\xi \over 2}\! \right )^{1/2} \!\!
             P^{|m|}_i(\xi),~~\xi={R^2-b^2\over R^2+b^2}, \\
\sigma^{|m|}_j &=& \left (\! {2|m|+2j+1\over 2\pi b^2}\! \right )
               \left ( {1-\xi \over 2}\right )^{3/2}\!\!
               P^{|m|}_i(\xi),
\end{eqnarray}
that yield \citep{AI78,H80}
\begin{equation}
D_j(m)=D_j(-m)= -{(2|m|+j)!\over 2b j!}.
\end{equation}
$P^{|m|}_i(\xi)$ are associated Legendre functions with $i=|m|+j$.
Clutton-Brock functions have a length scale $b$, which makes
them suitable for reproducing the potential and surface density of
soft-centered models. The choice of this parameter is an important
step in the calculation of normal modes. I will discuss on this issue
later in \S\ref{sec:modes-exp-disk}.

Equating (\ref{eq:expansion-V1}) and (\ref{eq:expansion-V1-config}),
multiplying both sides of the resulting equation by 
$\exp[-{\rm i}(l\theta_R+m\theta_{\phi})]$ and integrating over the
$\left (\theta_R,\theta_{\phi} \right )$-space, lead to 
(see also Kalnajs 1977 and Tremaine \& Weinberg 1984)  
\begin{eqnarray}
&{}& \!\!\! \sum_{j=0}^{\infty} b^{ml}_j(t) \Psi^{ml}_j(\textbf{\textit{J}})
  \!=\! \sum_{j=0}^{\infty} a^m_j(t) \tilde \Psi^{ml}_j(\textbf{\textit{J}}),
 \label{eq:relation-b-and-a} \\
&{}& \!\!\! \tilde \Psi^{ml}_j(\textbf{\textit{J}}) \!=\!
     {1\over \pi}\int\limits_{0}^{\pi}
\psi^{|m|}_j(R) \cos [l\theta_R+m(\theta_{\phi}-\phi)] d \theta _R,
 \label{eq:fourier-coeffs}
\end{eqnarray}
where $\tilde \Psi^{ml}_j$ are the Fourier coefficients of the basis
potential functions in the configuration space. The trial functions
used in the expansion of $V_1$ in the action-angle space are not
necessarily identical to $\tilde \Psi^{ml}_j$. However, subsequent
mathematical derivations are greatly simplified by setting
$\Psi^{ml}_j(\textbf{\textit{J}})= \tilde
\Psi^{ml}_j(\textbf{\textit{J}})$, which implies
$b^{ml}_j(t)=a^m_j(t)$. To build a relation between $d^{ml}_j(t)$
and $a^m_j(t)$, I use the fundamental equation
\begin{equation}
f_1(\Theta,\textbf{\textit{J}},t) d \textbf{\textit{J}} d \Theta
=\Sigma_1(R,\phi,t) R dR d \phi,
\label{eq:fundamental-equation}
\end{equation}
where $ d \textbf{\textit{J}} d \Theta$ is the volume of
an infinitesimal phase space element. On substituting
(\ref{eq:expansion-f1}) and (\ref{eq:expansion-sigma1-config})
in (\ref{eq:fundamental-equation}), multiplying both sides of
the resulting equation by $\psi^{|m|}_j(R)e^{-{\rm i}m\phi}$ and
integrating, one obtains
\begin{eqnarray}
a^m_j(t)&=& \frac
{4\pi^2}{D_j(m)}\sum_{l=-\infty}^{\infty}\sum_{p=0}^{\infty}
\Lambda^{ml}_{jp}d^{ml}_p(t),\label{eq:a-versus-d} \\
\Lambda^{ml}_{jp} &=& \int
\Psi^{ml}_j(\textbf{\textit{J}})
\Phi^{ml}_p(\textbf{\textit{J}}) d \textbf{\textit{J}},
\end{eqnarray}
which is inserted in (\ref{eq:expansion-V1}) to represent $V_1$
in terms of the amplitude functions $d^{ml}_j(t)$ as
\begin{equation}
V_1 \! =\!\!\!\!
\sum_{m,l,k=-\infty}^{\infty}\sum_{j,p=0}^{\infty}\frac
{4\pi^2}{D_j(m)}\Lambda^{mk}_{jp}\Psi^{ml}_j(\textbf{\textit{J}})
d^{mk}_p(t)
e^{{\rm i}\left(m\theta_\phi+l\theta_R \right)}.
\label{eq:expansion-V1-vs-dmlj}
\end{equation}

It would be computationally favorable to collect $d^{ml}_j(t)$
in a single vector $\textbf{\textit{z}}(t)=\{z_n(t)\}$ by defining
a map $(m,l,j)\rightarrow n$. In practice the infinite sums in
(\ref{eq:expansion-f1}) are truncated and approximated by finite
sums so that $-l_{\rm max}\le l \le l_{\rm max}$,
$-m_{\rm max}\le m\le m_{\rm max}$ and $0\le j\le
j_{\rm max}$. For $1\le n\le n_{\rm max}$,
a simple map between indices will be
\begin{eqnarray}
n &=& \left ( m+m_{\rm max}\right )
\left ( 2l_{\rm max}+1\right )(j_{\rm max}+1)
\nonumber \\
&{}& + \left (l+l_{\rm max}\right )(j_{\rm max}+1)+j+1,
\label{eq:map-mlj-to-i} \\
n_{\rm max} &=& \left ( 2m_{\rm max}+1\right )
      \left ( 2l_{\rm max}+1\right )
      (j_{\rm max}+1). \label{eq:maximum-n}
\end{eqnarray}
One can now use (\ref{eq:expansion-f1}) and
(\ref{eq:expansion-V1-vs-dmlj}) in (\ref{eq:CBE-perturbed}) and
apply the Petrov-Galerkin method to construct the weighted residual
form of the CBE. That is to multiply (\ref{eq:CBE-perturbed}) by
some weighting functions $W^{ml}_j(\Theta,\textbf{\textit{J}})$ and to
integrate the identity over the action-angle space. The outcome is
the following system of nonlinear ODEs 
\begin{equation}
{\rm i} { d z_p\over  d t}\!=\! \sum_{q=1}^{n_{\rm max}}
\! A_{pq} z_q \!+ \!\! \sum_{q,r=1}^{n_{\rm max}}
\! B_{pqr} z_q z_r,~~p=1,2,\cdots,n_{\rm max},
\label{eq:nonlinear-ODE}
\end{equation}
for the amplitude functions $z_n(t) \equiv d^{ml}_j(t)$. The
elements of $A_{pq}$ and $B_{pqr}$ have been determined in Appendix
A. Each equation in (\ref{eq:nonlinear-ODE}) is the projection of
the CBE on a subspace spanned by a weighting function. 
Therefore, the left
hand side of (\ref{eq:nonlinear-ODE}) is the projection of $\partial
f_1/\partial t$, the summation over first order terms is the
projection of $-\left [f_1,{\cal H}_0 \right ]- \left [ f_0,{\cal
H}_1 \right ]$, and the second order terms are the projections of
$-\left [f_1,{\cal H}_1 \right ]$. The second order terms of
amplitude functions, characterized by $B_{pqr}$, show the
interaction of modes in both the radial and azimuthal directions.  

Distribution of angular momentum between different Fourier components
provides useful information of the disk dynamics. I compute the rate
of change of the total angular momentum ${\cal L}$ using
(see Appendix B in JH)
\begin{equation}
{ d {\cal L}\over  d t}=-\frac {1}{4}
\int\int \left ( f_1+\overline{f}_1 \right )
{\partial \over \partial \theta_{\phi} }
\left ( V_1+\overline{V}_1 \right ) d \textbf{\textit{J}} d \Theta,
\label{eq:rate-change-L-one}
\end{equation}
where a bar denotes complex conjugate. Substituting
(\ref{eq:expansion-f1}) and (\ref{eq:expansion-V1-vs-dmlj})
in (\ref{eq:rate-change-L-one}) and evaluating the integrals,
yield
\begin{eqnarray}
{ d {\cal L}\over  d t} &=& {\rm i}\pi^2 \!\!\!
\sum_{m,l=-\infty}^{\infty}\sum_{j,p=0}^{\infty}
\biggl \{  m\left [ a^{(-m)}_{j}(t)
+\overline{a^{m}_{j}}(t) \right ] \Lambda^{ml}_{jp}
d^{ml}_{p}(t) \nonumber \\
&{}& - m\left [ a^{m}_{j}(t)
+\overline{a^{(-m)}_{j}}(t) \right ] \Lambda^{ml}_{jp}
\overline{d^{ml}_{p}}(t) \biggr \}. \label{eq:rate-change-L-two}
\end{eqnarray}
Define $a^m_j(t)=u^m_j(t)+{\rm i}v^m_j(t)$ with $u^m_j(t)$ and
$v^m_j(t)$ being real functions of time. According to identity
(\ref{eq:a-versus-d}), one may further simplify equation
(\ref{eq:rate-change-L-two}) to
\begin{eqnarray}
{ d {\cal L}\over  d t} &=&
\sum_{m=-\infty}^{\infty} L_m(t),
\label{eq:rate-change-L-three} \\
L_m(t) &=& -\frac {m}{2}\sum_{j=0}^{\infty}
D_j(m) \nonumber \\
&{}& \times \left [ u^m_j(t)v^{-m}_j(t)+u^{-m}_j(t)v^m_j(t) \right ].
\label{eq:rate-change-L-four}
\end{eqnarray}
The share of the $m$th mode from $ d {\cal L}/ d t$ is thus
determined by $L_m(t)$. As one could anticipate for an isolated
stellar disk, $ d {\cal L}/ d t$ vanishes and the total
angular momentum remains constant because the terms $L_m(t)$ and
$L_{-m}(t)$ cancel each other in (\ref{eq:rate-change-L-three}),
and $L_0(t)$ is annulled by the factor $m$
in (\ref{eq:rate-change-L-four}).

\subsection{Trial and Weighting Functions}
\label{sec::trial-and-test-fncs}

Choosing the trial functions $\Phi^{ml}_j(\textbf{\textit{J}})$ is
the most delicate step in the reduction of the CBE to a system of ODEs.
One possible way is to set $\partial f_1/\partial t=0$ in
(\ref{eq:CBE-perturbed}) and solve the first order equation
\begin{equation}
\left [f_1,{\cal H}_0 \right ]+\left [f_0,{\cal H}_1 \right ]=0,
\label{eq:CBE-linear-equilibrium}
\end{equation}
for $f_1$. Substituting (\ref{eq:expansion-f1}) and
(\ref{eq:expansion-V1}) in (\ref{eq:CBE-linear-equilibrium}) gives
\begin{eqnarray}
d^{ml}_j\Phi^{ml}_j(\textbf{\textit{J}}) &=& b^{ml}_j
\varrho^{ml}_{0}(\textbf{\textit{J}}) \Psi^{ml}_j(\textbf{\textit{J}}),
\label{eq:trial-functions-one} \\
\varrho^{ml}_{0}(\textbf{\textit{J}}) &=& {l\frac {\partial f_0}{\partial
J_R}+ m\frac {\partial f_0}{\partial J_\phi} \over
l\Omega_R+m\Omega_\phi },
\end{eqnarray}
where
\begin{equation}
\Omega_R(\textbf{\textit{J}})=\frac {\partial {\cal H}_0}{\partial J_R},~~
\Omega_\phi(\textbf{\textit{J}})=\frac {\partial {\cal H}_0}{\partial J_\phi}.
\end{equation}
Equation (\ref{eq:trial-functions-one}) suggests to choose
\begin{equation}
\Phi^{ml}_j(\textbf{\textit{J}})=\varrho^{ml}_{0}(\textbf{\textit{J}})
\Psi^{ml}_j(\textbf{\textit{J}}),\label{eq:trial-functions-two}
\end{equation}
as the trial functions (in the space of actions) for an unsteady
$f_1(\Theta,\textbf{\textit{J}},t)$. These functions have integrable
singularities for resonant orbits with $l\Omega_R+m\Omega_{\phi}=0$.
For unidirectional disks with only prograde orbits, they also
include a term with the Dirac delta function $\delta(J_\phi)$
(see JH). One should therefore avoid the partial derivatives of
$\Phi^{ml}_j$ with respect to the actions by evaluating the weighted
residual form of $[f_1,{\cal H}_1]$ through integration by parts
(Appendix A).
 
For deriving the relation between $a^m_j$ and $d^{ml}_{j}$
in (\ref{eq:a-versus-d}), the fundamental equation
(\ref{eq:fundamental-equation}) was multiplied by the complex
conjugates of the basis functions used in the expansion of
$V_1(R,\phi,t)$. One may follow a similar approach for
obtaining the weighted residual form of the CBE and set
\begin{equation}
W^{ml}_j(\Theta,\textbf{\textit{J}})=\Psi^{ml}_j(\textbf{\textit{J}})
e^{-{\rm i}(m\theta_\phi+l\theta_R)},\label{eq:weight-functions}
\end{equation}
which are the complex conjugates of the basis functions used in the 
expansion of $V_1(\Theta,\textbf{\textit{J}},t)$ in the action-angle 
space. The trial and weighting functions introduced as above, are not 
orthogonal but they result in a simple form for the linear part of 
the reduced CBE as I explain in \S\ref{sec::linear-theory}.

\section{LINEAR THEORY}
\label{sec::linear-theory}

In a first order perturbation analysis, the second order terms of
the amplitude functions are ignored. The evolution of modes is
then governed by the linear parts of (\ref{eq:ODE-for-z-one}) as
\begin{equation}
{\rm i} \textbf{\textit{M}}\cdot { d \over  d t}
\textbf{\textit{z}}(t) =
\textbf{\textit{C}}\cdot \textbf{\textit{z}}(t).
\label{eq:ODE-for-z-linear}
\end{equation}
A general solution of (\ref{eq:ODE-for-z-linear}) has the
form $\textbf{\textit{z}}(t)=e^{-{\rm i}\omega t}\textbf{\textit{z}}_0$,
which leads to the following linear eigenvalue problem
\begin{equation}
\textbf{\textit{C}}(m)\cdot \textbf{\textit{z}}_0=
\omega \textbf{\textit{M}}(m)\cdot \textbf{\textit{z}}_0,
\label{eq:eigenvalue-problem-one}
\end{equation}
for a prescribed azimuthal wavenumber $m$.
Operating $\textbf{\textit{M}}^{-1}$ on (\ref{eq:eigenvalue-problem-one})
yields
\begin{equation}
\textbf{\textit{A}}(m)\cdot \textbf{\textit{z}}_0=
\omega \textbf{\textit{z}}_0,
\label{eq:eigenvalue-problem-two}
\end{equation}
where $\textbf{\textit{A}}$ is a general non-symmetric matrix. A reduction
to Hessenberg form followed by the QR algorithm \citep{Press01}
gives all real and complex eigenvalues. Real eigenvalues correspond
to van Kampen modes and complex eigenvalues, which occur in conjugate
pairs, give growing/damping modes. I utilize the method of singular value
decomposition for finding the eigenvectors and perform the decomposition
$\textbf{\textit{A}}-\omega \textbf{\textit{I}}=\textbf{\textit{U}}\cdot
\textbf{\textit{S}}\cdot \textbf{\textit{V}}^T$
where $\textbf{\textit{I}}$ is the identity matrix and the diagonal matrix
$\textbf{\textit{S}}$ is composed of the singular values $S_j$
($j=1,2,\cdots,n_{\rm max}$). The column of $\textbf{\textit{V}}$ that
corresponds to the smallest $S_j$ is the eigenvector associated
with $\omega$.

Calculation of $\textbf{\textit{C}}$ and $\textbf{\textit{M}}$
involves evaluation of some definite integrals in the action space.
There will be two types of such integrals (instead of three) if one
uses the trial functions defined in (\ref{eq:trial-functions-two}).
Let me introduce the auxiliary integral
\begin{equation}
{\cal I}^{ml}_{jk}=
\int  d \textbf{\textit{J}} \left ( l{\partial f_0\over \partial J_R}+
m{\partial f_0\over \partial J_\phi} \right )
\Psi^{ml}_{j}(\textbf{\textit{J}})\Psi^{ml}_{k}(\textbf{\textit{J}}),
\end{equation}
and apply the trial functions $\Phi^{ml}_{j}=\varrho^{ml}_0\Psi^{ml}_{j}$
in (\ref{eq:A-tensor}). The elements of $\textbf{\textit{M}}$ and
$\textbf{\textit{C}}$ are thus computed from
\begin{eqnarray}
M_{pq} &=& \delta_{l,l'}\Lambda^{ml}_{jj'}, \label{eq:bar-M-matrix} \\
C_{pq} &=& \delta_{l,l'}{\cal I}^{ml}_{jj'}-
\sum_{k=0}^{j_{\rm max}} \left [ {4\pi^2\over D_{k}(m)}\right ]
{\cal I}^{ml}_{jk}\Lambda^{ml'}_{kj'}. \label{eq:bar-A-matrix}
\end{eqnarray}
Both $\Lambda^{ml}_{jk}$ and ${\cal I}^{ml}_{jk}$ consist of
boundary integrals when the unperturbed stellar disk is
unidirectional with the DF
$f_0(\textbf{\textit{J}})=H(J_\phi)f^P_0(\textbf{\textit{J}})$.
Here $H$ is the Heaviside function. The boundary terms are
\begin{eqnarray}
\tilde \Lambda^{ml}_{jk} &=& \int_{0}^{\infty} d J_R\left [
{mf^P_0(\textbf{\textit{J}})
\Psi^{ml}_{j}(\textbf{\textit{J}})\Psi^{ml}_{k}(\textbf{\textit{J}}) \over
l\Omega_R(\textbf{\textit{J}})+m\Omega_{\phi}(\textbf{\textit{J}}) }
\right ]_{J_{\phi}=0}, \\
{\tilde {\cal I}}^{ml}_{jk} &=& \int_{0}^{\infty} d J_R\left [
mf^P_0(\textbf{\textit{J}})
\Psi^{ml}_{j}(\textbf{\textit{J}})\Psi^{ml}_{k}(\textbf{\textit{J}}) \right
]_{J_{\phi}=0}.
\end{eqnarray}

Dynamics of modes with different azimuthal wavenumbers are decoupled
in the linear regime and the matrix $\textbf{\textit{A}}$ is an odd
function of the wavenumber $m$. i.e.,
$\textbf{\textit{A}}(-m)=-\textbf{\textit{A}}(m)$. An
immediate result of this property is $a^{-m}_j(t)=\overline {a^m_j}(t)$.
Consequently, $L_m(t)$ becomes equal to zero for all $|m|\ge 0$ and
each mode individually conserves the total angular momentum.

The present theory has three major advantages over Kalnajs's formulation.
Firstly, all eigenmodes relevant to a prescribed azimuthal wavenumber
are obtained at once with classical linear algebraic algorithms. This makes
it possible to explore and classify all families of growing modes beside
pure oscillatory van Kampen modes. Secondly, the constituting integrals
of the elements of $M_{pq}$, $C_{pq}$ and $K_{pqr}$
(Appendix \ref{app::Petrov-Galerkin}) are regular
at exact resonances when the condition $l\Omega_R+m\Omega_\phi-\omega=0$
holds. Finally, nonlinear interaction of modes, and the mass and angular
momentum exchange between them, can be readily monitored by integrating
the system of nonlinear ODEs given in (\ref{eq:nonlinear-ODE}). In the
proceeding section I will be concerned with the calculation and
classification of modes in the linear regime.

\begin{figure*}
\plottwo{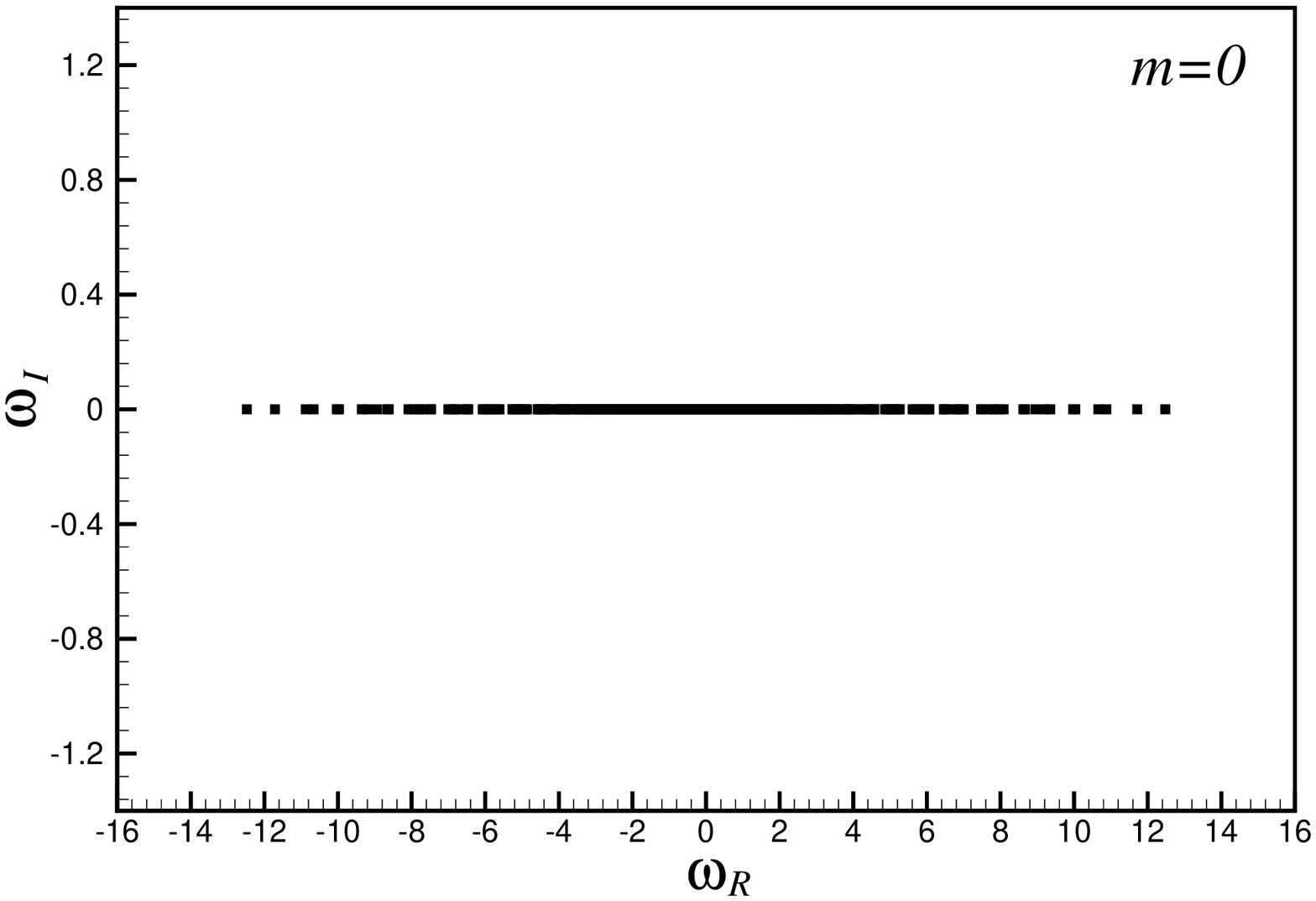}{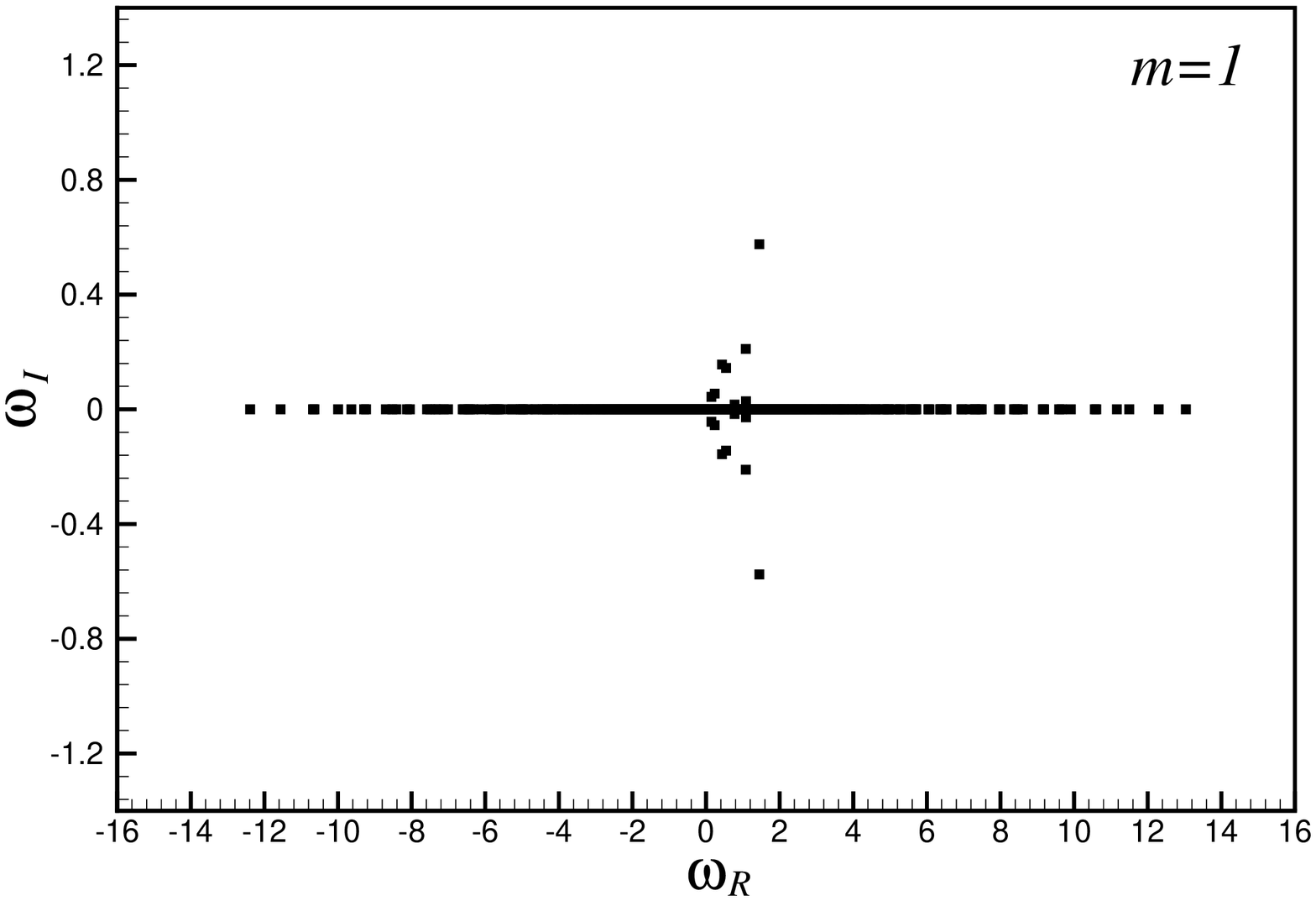}
\plottwo{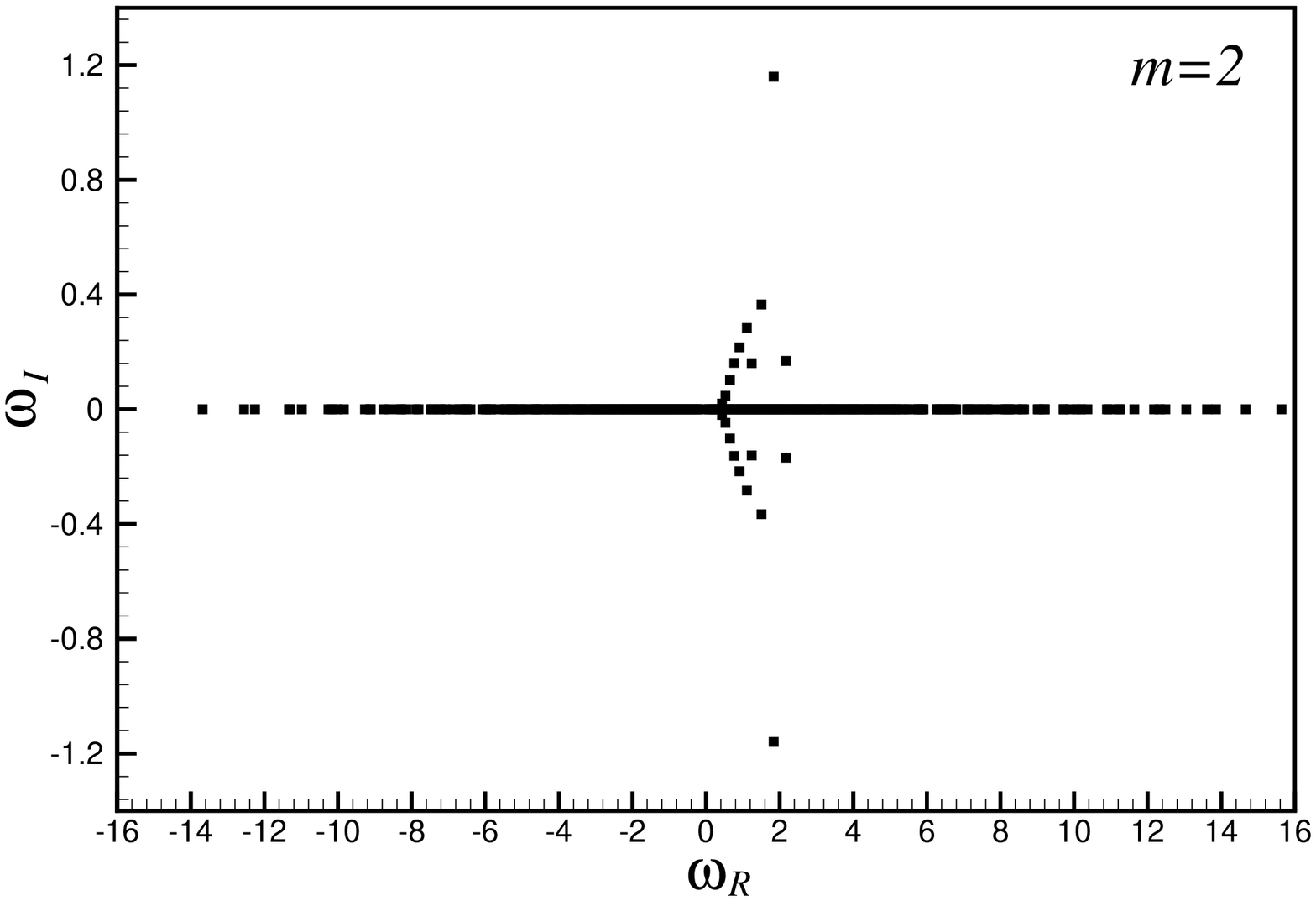}{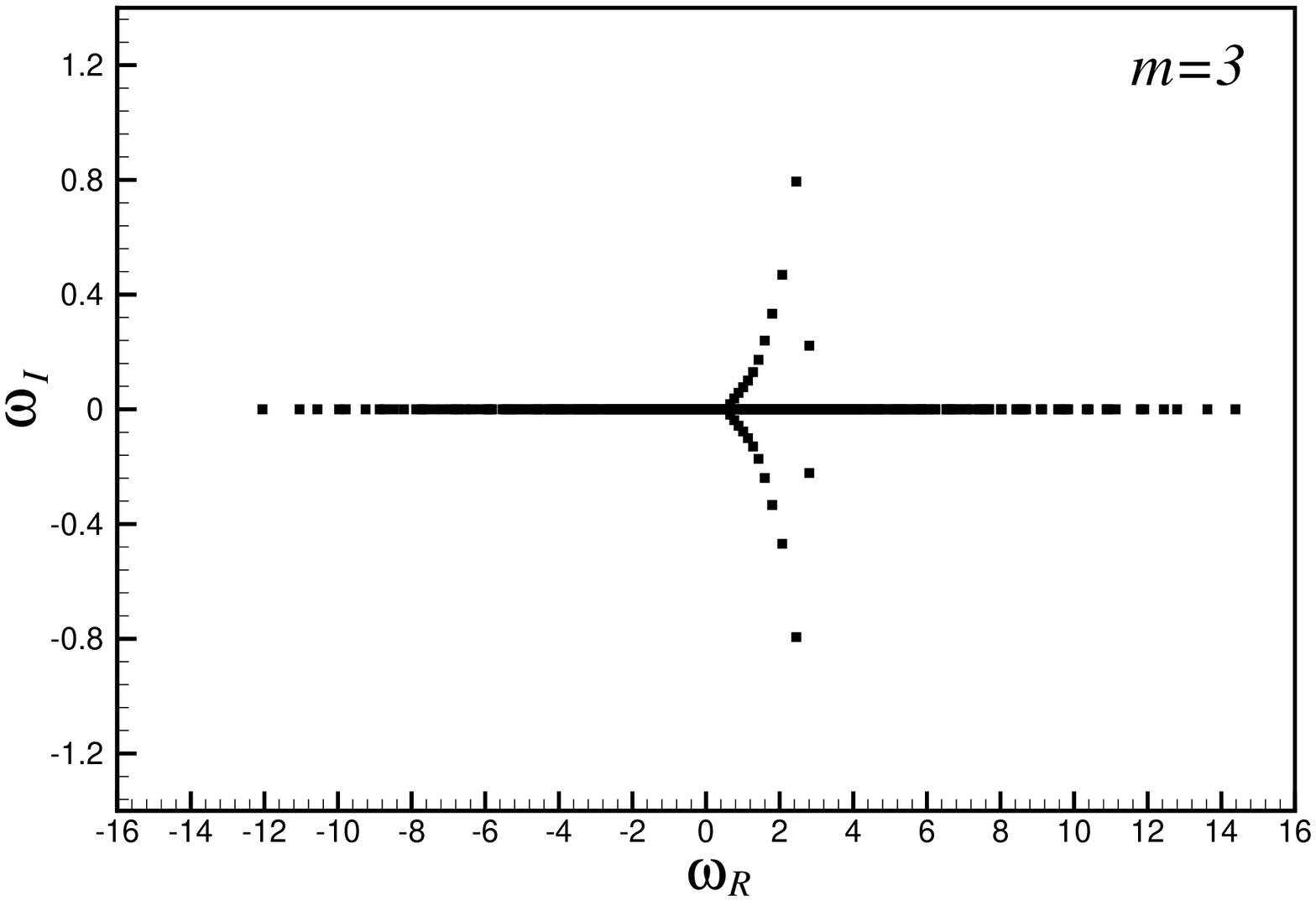}
\plottwo{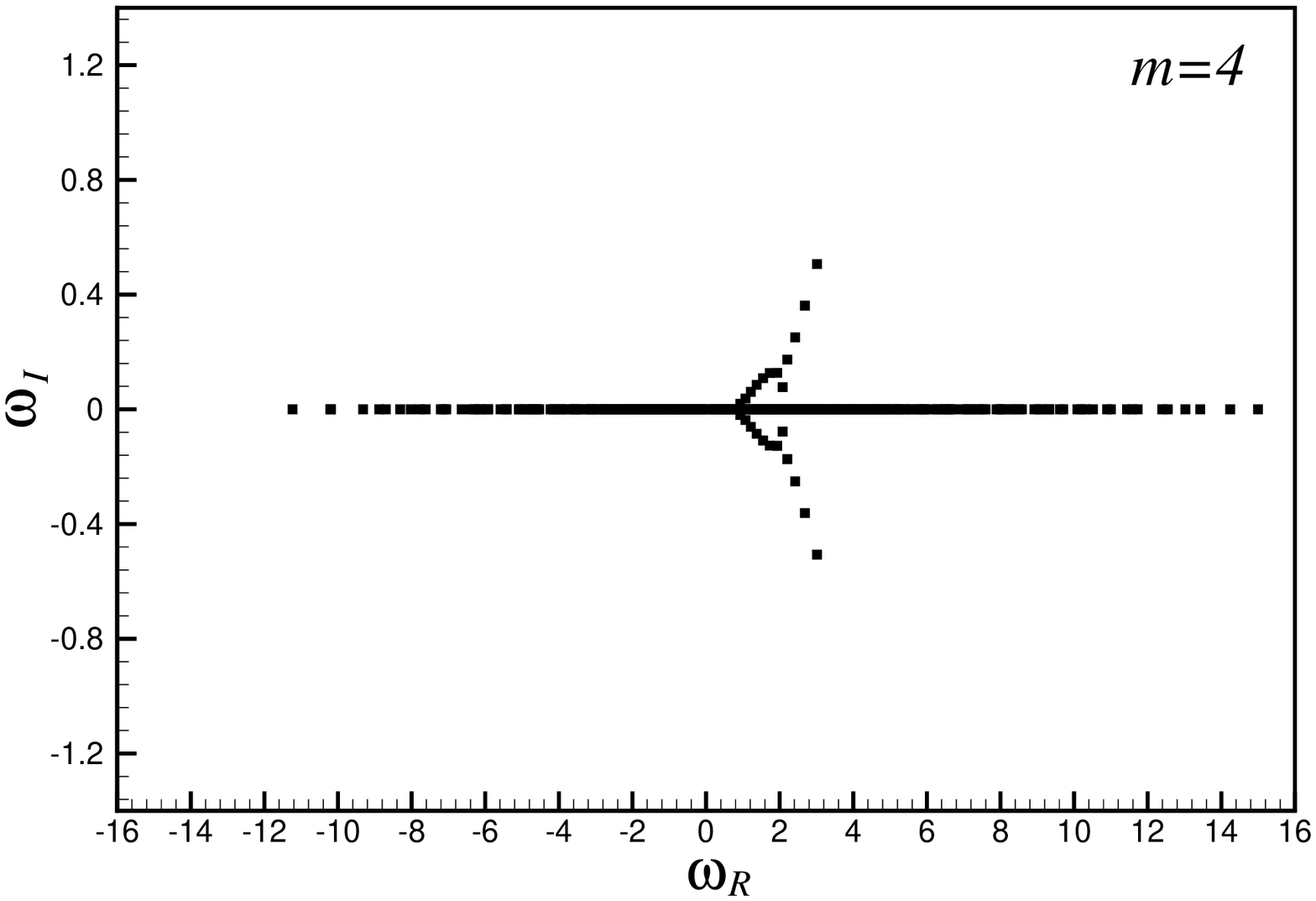}{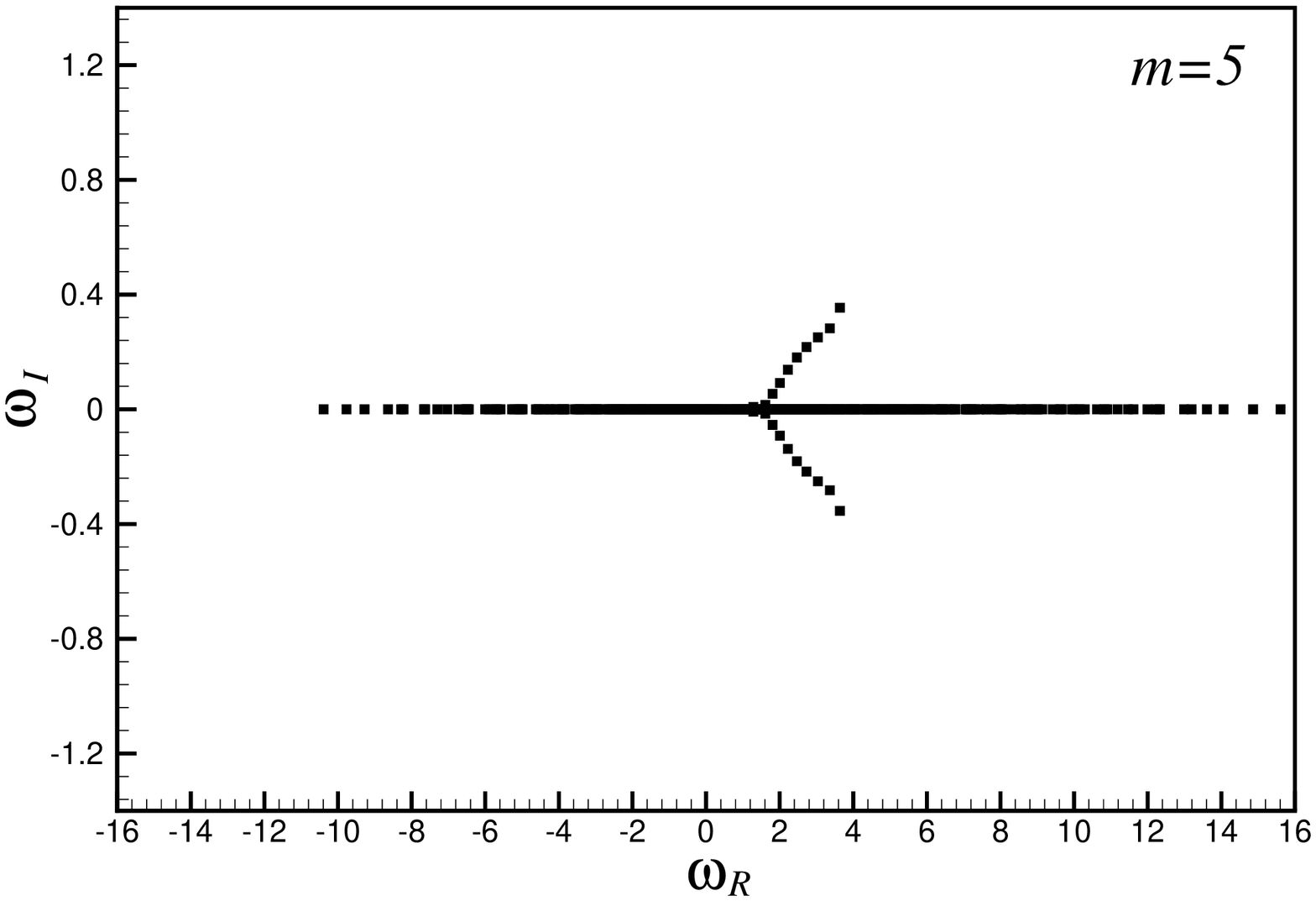}
\caption{Eigenfrequency spectra of a cored exponential disk
with $v_0=1$, $R_C=1$ and $(N,\lambda,\alpha)=(6,1,0.42)$.
Eigenfrequencies have been displayed for the azimuthal wavenumbers
$0\le m \le 5$. The results correspond to $l_{\rm max}=10$ and
$j_{\rm max}=15$ in the series expansion of the perturbed
distribution function. \label{pic:full-spectra}}
\end{figure*}
\begin{figure*}
\plottwo{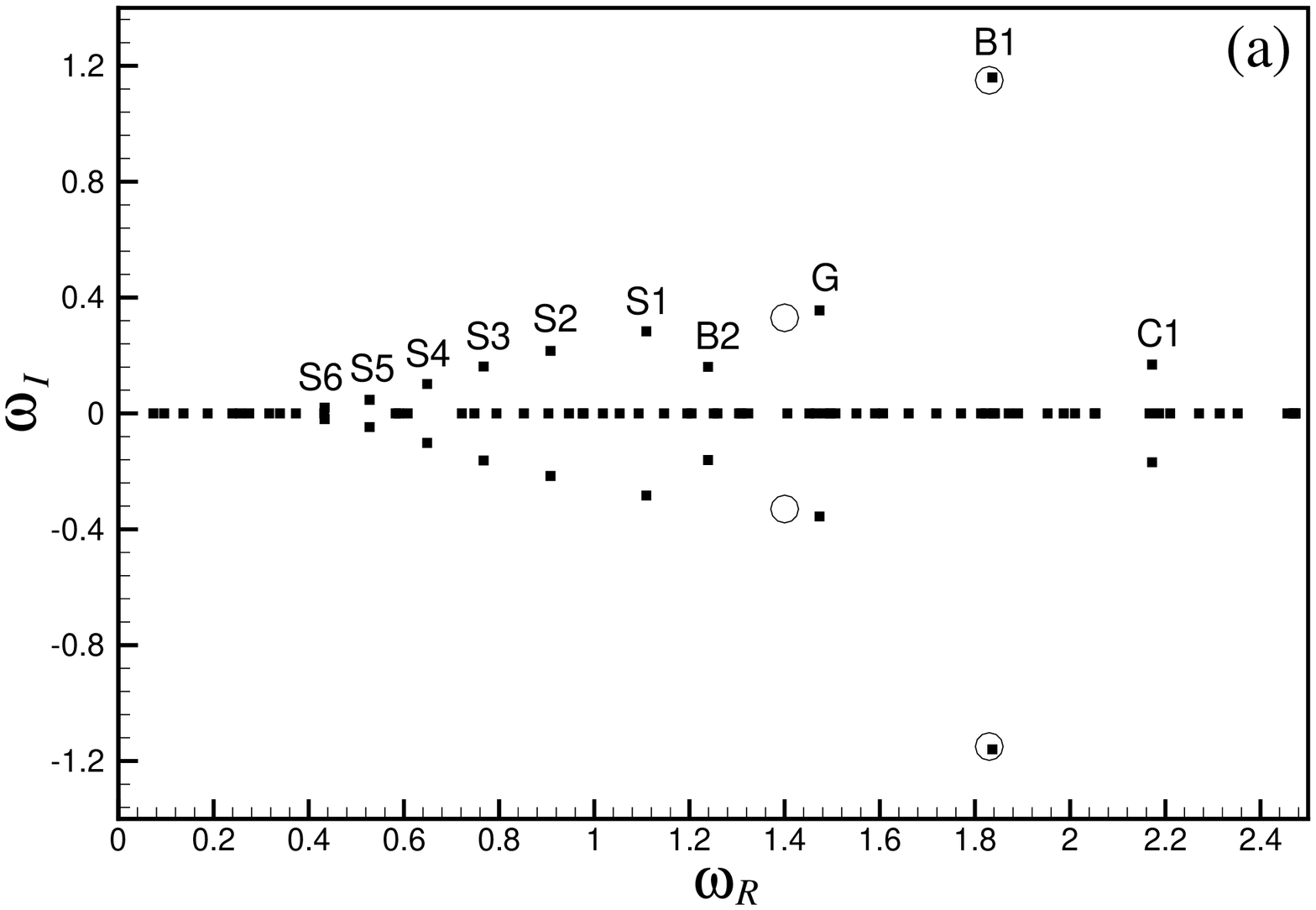}{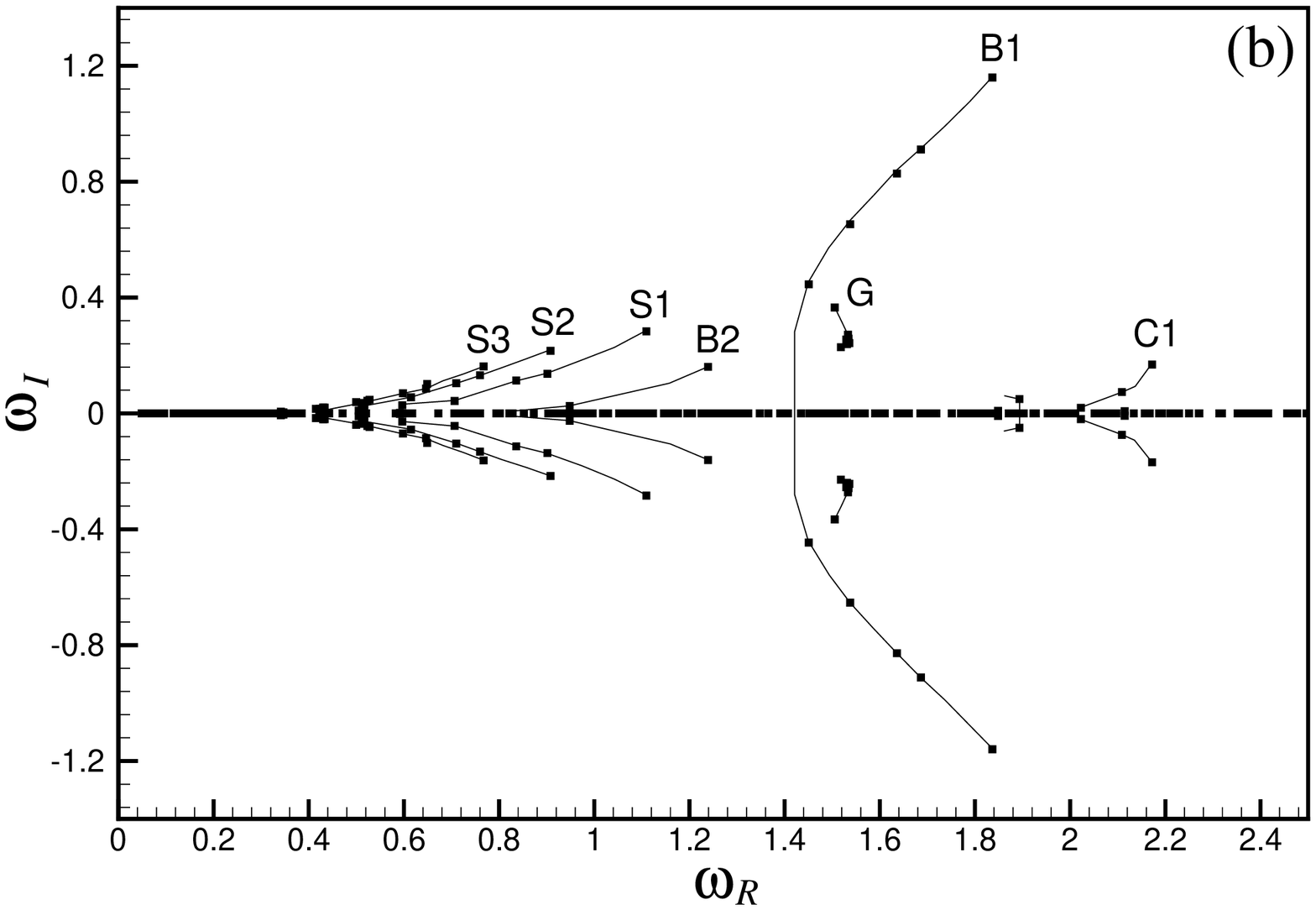}
 \caption{({\em a}) Zoomed
eigenfrequency spectra (filled squares) of a cored exponential disk 
with $(N,\lambda,\alpha)=(6,1,0.42)$ for the azimuthal wavenumber $m=2$.
The (isolated) bar mode has been labeled B1. The most prominent
growing modes belong to a discrete family that bifurcates from a van
Kampen mode. The members of this family, labeled as S1, S2,
S3,$\cdots$, have spiral patterns. Circles show the eigenfrequencies
of the fundamental and secondary modes calculated using Kalnajs's
(1977) method (see Table 4 in JH). ({\em b}) The eigenfrequency loci
of the same model of panel {\em a} as $\alpha$ is increased
continuously from $0.2$ to $0.42$. A few sample eigenfrequencies
have been displayed on each locus.
 \label{pic:spectrum-m2-unstable}} 
\end{figure*}
\begin{figure*}
\plotone{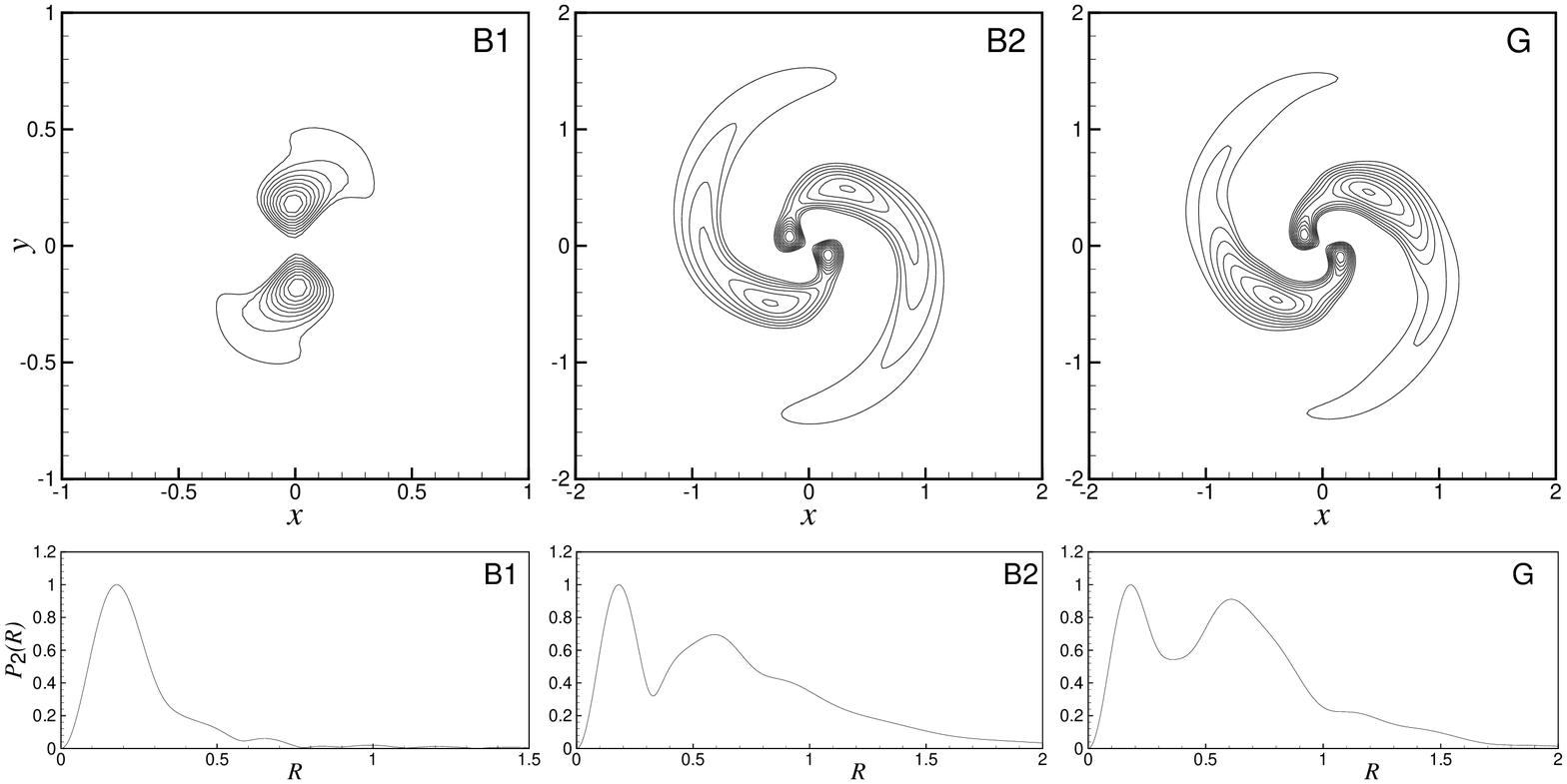}
\plotone{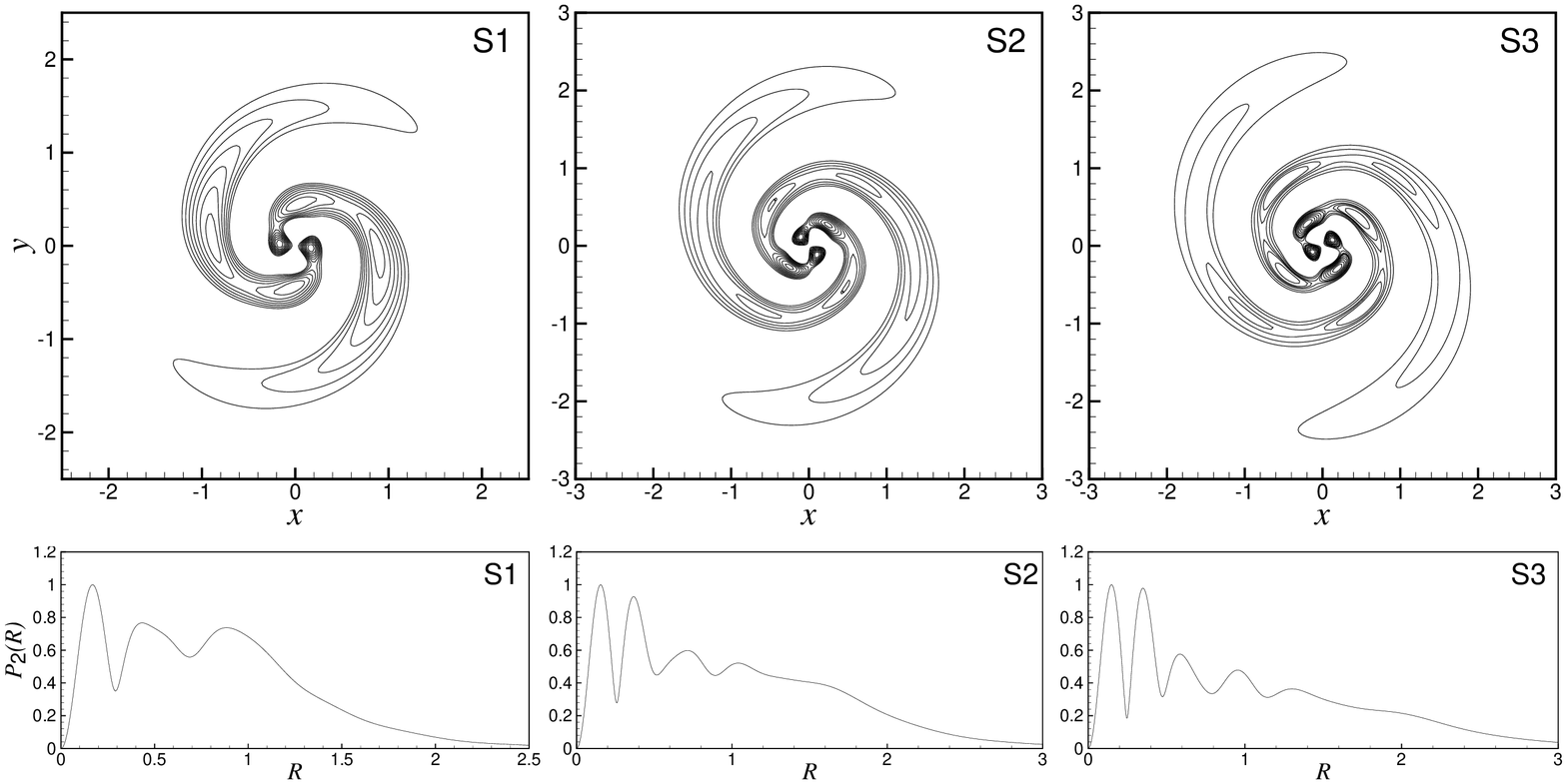}
\caption{Mode shapes of the cored exponential disk for $R_C=1$,
$v_0=1$ and $(N,\lambda,\alpha)=(6,1,0.42)$. Panels have been
labeled by the corresponding mode name. The contour plots show the
positive part of $\Sigma_1(R,\phi,0)$. The contour levels range
from 10$\%$ to 90$\%$ of the maximum of $\Sigma_1(R,\phi,0)$ with
increments of 10$\%$. The panel below each mode shape shows the
amplitude of wave patterns as defined in equation (\ref{eq:mode-shape}).
\label{pic:mode-shape-m2}}
\end{figure*}

\section{MODES OF THE CORED EXPONENTIAL DISK}
\label{sec:modes-exp-disk}

JH calculated barred and spiral modes of certain stellar
disks for the wavenumber $m=2$. Among the models
studied in JH, the cored exponential disk with the surface
density profile
\begin{equation}
\Sigma_D(R)=\Sigma_s \exp \left ( -\lambda
\sqrt{1+R^2/R_C^2}\right ),~~\lambda=\frac{R_C}{R_D},
\label{eq:density-exp-disk}
\end{equation}
and embedded in the field of the soft-centered logarithmic
potential
\begin{equation}
V_0(R)=v_0^2\ln \sqrt{1+R^2/R_C^2},
\label{eq:potential-cored-logarithmic}
\end{equation}
is a viable model that resembles most features of realistic spirals.
Here $R_C$ is the core radius, $R_D$ is the length scale of
the exponential decay, and $\Sigma_s$ is a density scaling factor.
The velocity of circular orbits in this model rises from zero at
the galactic center and approaches to the constant value $v_0$
in outer regions where the light profile falls off exponentially.
\citet{JHb} have derived the gravitational potential corresponding
to $\Sigma_D(R)$. I denote this potential by $V_D(R)$. The gradient
\begin{equation}
F_H=\frac{d}{dR}\left [V_0(R)-V_D(R)\right ],
\end{equation}
will give the gravitational force of a spherical dark matter component,
computed inside the galactic disk. The density profile of the dark
component, $\rho_H$, can then be determined using $F_H$. The positiveness
of $\rho_H$ imposes some restrictions on the physical values of $\lambda$
and $\alpha=G\Sigma_s R_D/v_0^2$ as Figure 5 in JH shows. For a given
$\lambda$, $\alpha$ cannot exceed a critical value $\alpha_{cr}$.
The parameter $\lambda$ determines the shape of the dark matter density 
profile. A model with $\lambda=1$ and $\alpha=\alpha_{cr}$ is maximal 
in the region where the rotation curve is rising. i.e., there is no
dark matter in that region. Models with $\lambda >1$ and
$\alpha=\alpha_{cr}$ are still maximal but only in the vicinity of
the center for $R<R_D$. In such models the rotational velocity of
stars due to dark matter ($v_H=\sqrt{R F_H}$) has a monotonically
rising profile. For $\lambda<1$, dark matter penetrates into the 
galactic center and its density profile becomes cuspy in the limit 
of $\lambda \rightarrow 0$. The role of the parameter $\alpha$ is 
to control the fraction of dark to luminous matter. Models with 
$\alpha \ll \alpha_{cr}$ are dominated by dark matter. 
 
JH introduced a family of equilibrium DFs that reproduces
$\Sigma_D(R)$ and depends on an integer constant $N$.
This parameter controls the population of near-circular
orbits and the disk temperature: the parameter $Q$ of \citet{T64}
decreases by increasing $N$. The DFs of JH have an isotropic part
that determines the fraction of radial orbits. That isotropic part,
which reconstructs the central density of the equilibrium state,
shrinks to central regions of the galaxy as $N$ increases.

I apply my new method to the cored exponential disks of JH and calculate
the spectrum of $\omega=\omega_R+{\rm i}\omega_I$. Subsequently, 
the eigenvector $\textbf{\textit{z}}_0$ is calculated 
from (\ref{eq:eigenvalue-problem-two}) and it is used 
in (\ref{eq:a-versus-d}) to compute
$a^m_j(t)=e^{-{\rm i}\omega t}a^m_j(0)$ and the perturbed density
\begin{equation}
\Sigma_1(R,\phi,t)= e^{\omega_I t} P_m(R) \cos \left [
m\phi \!-\! \omega_R t \!+\! \vartheta_m(R)\right ],
\label{eq:mode-shape}
\end{equation}
which is the real part of (\ref{eq:expansion-sigma1-config}).
$P_m(R)$ and $\vartheta_m(R)$ are the amplitude and phase functions
of an $m$-fold circumferential wave that travels with the angular
velocity $\omega_R/m$. The factor $e^{\omega_I t}$ shows the exponential
growth/decay of the wave amplitude. I normalize all length, velocity, 
and time variables to $R_C$, $v_0$ and $R_C/v_0$, respectively, and 
set $G=R_C=v_0=1$.  

I begin my case studies in \S\ref{sec:maximal-disk} with a near 
maximal disk of $(N,\lambda,\alpha)$=$(6,1,0.42)$ and compute its 
eigenfrequency spectra for the wavenumbers $0\le m\le 5$. I then 
classify unstable $m=2$ modes of this model and investigate their 
evolution as the parameter $\alpha$ is varied. 
In \S\ref{sec:variations-of-lambda} and \S\ref{sec:variation-of-N},
I study the behavior of unstable $m=2$ waves as the parameters 
$\lambda$ and $N$ are changed. The eigenfrequency spectrum of a 
model with an inner cutout is also computed and discussed 
in \S\ref{sec:cutout-models}.

\subsection{A Near Maximal Disk}
\label{sec:maximal-disk}

I pick up the first model from Table 4 of JH with
$(N,\lambda,\alpha)=(6,1,0.42)$ and start solving the
eigensystem (\ref{eq:eigenvalue-problem-two})
with $(l_{\rm max},j_{\rm max})=(2,4)$ and increase these limits
until complex eigenfrequencies converge. For an error threshold
of $1\%$ the program terminates when
$(l_{\rm max},j_{\rm max})=(10,15)$, which gives a size of
$336\times 336$ for the matrix $\textbf{\textit{A}}$. In such a
circumstance, out of $336$ eigenfrequencies of $\textbf{\textit{A}}$
(for each wavenumber $m$), less than 15 pair have non-zero
growth rates ($\omega_I\not =0$). Further increasing of
$l_{\rm max}$ and $j_{\rm max}$ does not alter the number and
location of complex eigenfrequencies in the $\omega$-plane.
This shows that unstable modes do not constitute a continuous
family.

Figure \ref{pic:full-spectra} displays the eigenfrequency spectra
for the azimuthal wavenumbers $0\le m\le 5$. Eigenfrequencies on
the real axis are oscillatory van Kampen modes. Their calculation
requires evaluation of Cauchy's principal value \citep{V03}
if one uses Kalnajs's first order theory. In the present formalism,
van Kampen modes are found together with growing modes without any
special treatment. More van Kampen modes are obtainable
by increasing the truncation limit $l_{\rm max}$ of Fourier terms
in the $\theta_R$-direction. Toomre's $Q$ is marginally greater
than 1 for the model (Figure 7{\em b} in JH), and therefore,
one could expect that the disk is stable for $m=0$ excitations
(see top-left panel in Figure \ref{pic:full-spectra}).
The model is highly unstable for $m>0$ excitations although the
average growth rate of unstable modes decreases for larger
wavenumbers. It is evident that either unstable modes are isolated
or they are grouped in {\it discrete families}. 
Depending on the wavenumber, there may be one or more discrete 
families. The most prominent family bifurcates from van Kampen modes. 
Members of this family have spiral patterns with multiple peaks in 
their $P_m(R)$ functions. The (global) fastest growing mode belongs 
to the spectrum of $m=2$. That is the bar mode of a two-member 
unstable family. 

The length scale of Clutton-Brock functions has been set to $b=1.5$
for $m=2$ and $b=2$ for other wavenumbers. Changing this length
scale slightly displaces the eigenfrequencies although the spectrum
maintains its global pattern. Large values of $b$ lead to a better
computation accuracy of extensive modes (with smaller pattern
speeds), while compact bar modes show a rapid convergence for small
values of $b$. Moreover, the suitable value of $b$ differs from one
azimuthal wavenumber to another. Finding an optimum length scale
that gives the best results for all modes and wavenumbers is an open
problem yet to be investigated precisely. For the cored exponential 
disks with $0.5 \le R_C,R_D\le 2$, working in the range $1\le b\le 2.5$ 
gives reasonable results.

\begin{figure*}
\plottwo{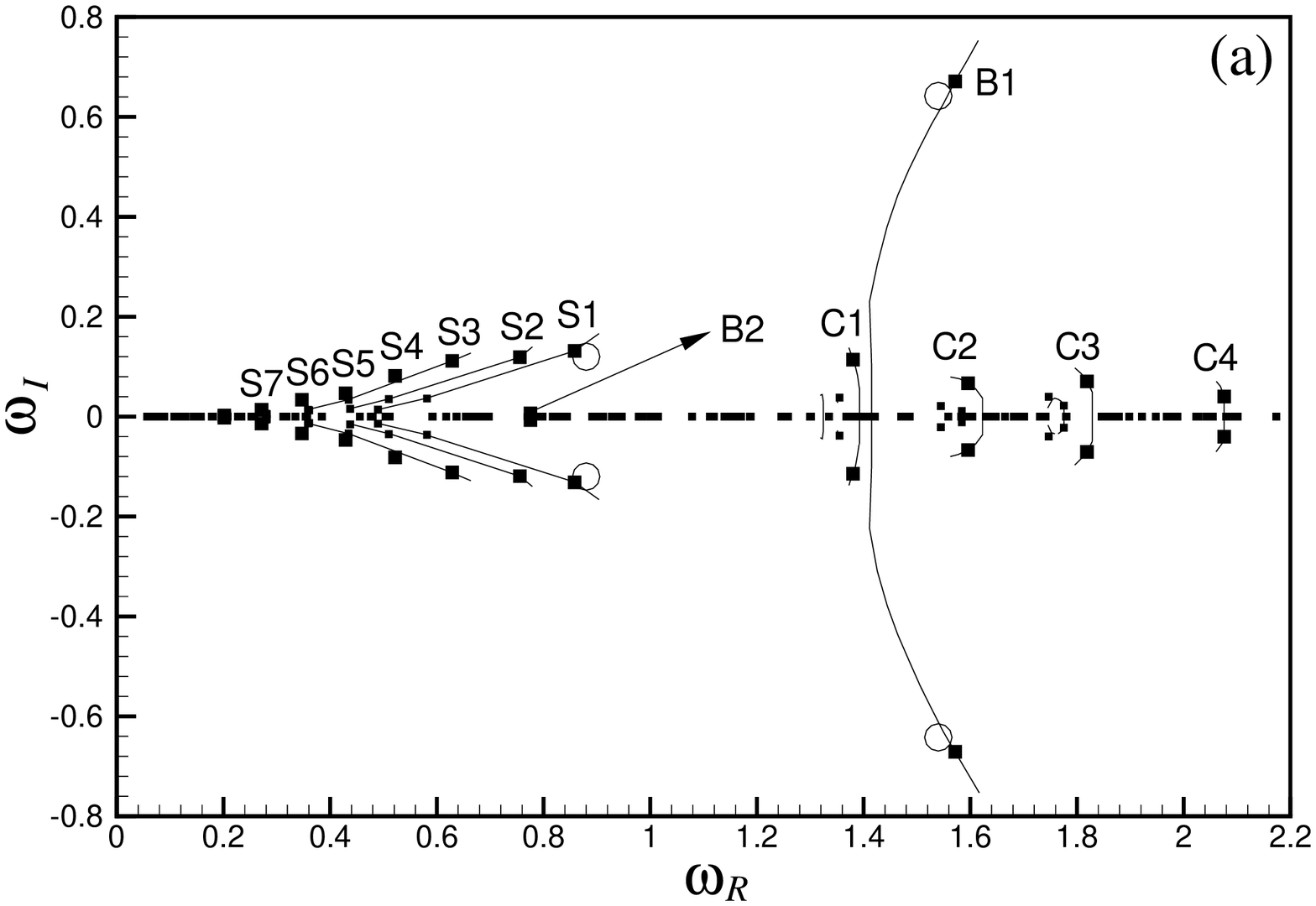}{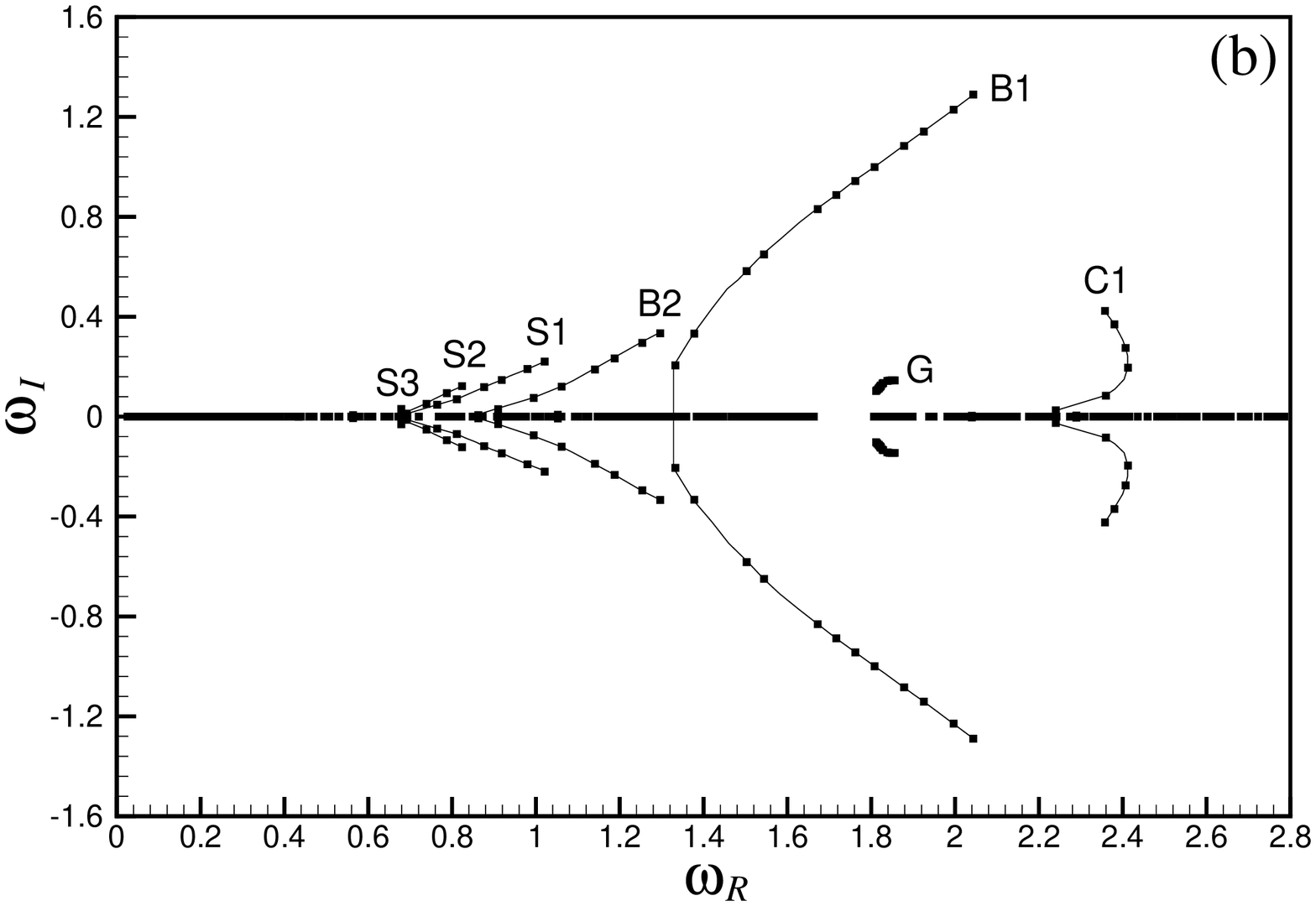} 
\caption{({\em a}) Eigenfrequency loci (solid lines) of cored 
exponential disks with $(N,\lambda)=(6,0.625)$ for $0.2\le \alpha \le 0.36$. 
Large squares show the eigenfrequency spectrum of a model 
with $\alpha=0.34$, and circles show the eigenfrequencies of the 
fundamental and secondary modes calculated using Kalnajs's method 
(see Table 4 in JH) for $\alpha=0.34$. ({\em b}) Same as panel {\em a} 
but for models with $(N,\lambda)=(6,2)$ and $0.25 \le \alpha \le 0.6$.
Some sample eigenfrequencies have been demonstrated on each locus.
\label{pic:spectrum-m2-vary-lambda}}
\end{figure*}

For $m=2$, I have zoomed out and plotted in Figure
\ref{pic:spectrum-m2-unstable}{\em a} the portion of the spectrum
that contains growing modes. The first and second modes reported in
Table 4 of JH have been shown by circles in the same figure. The
most unstable mode (labeled as B1) is a compact, rapidly rotating
bar. The majority of unstable modes belong to a discrete spiral
family that bifurcates from a van Kampen mode with $\omega \approx
0.43$. I have labeled these modes by S1,$\cdots$,S6. The number of
density peaks along the spiral arms is proportional to the integer
number in the mode name. Both B2 and G are double peaked spirals but
I have classified B2 as a bar mode, and collected it with B1 in a
two member family, for it takes a bar-like structure when it is 
stabilized by decreasing $\alpha$. I classify mode G as an isolated 
mode because it does not behave similar to either of S- or B-modes 
as the model parameters vary. There is another isolated mode in 
the spectrum, C1, which exhibits a spiral pattern. By decreasing
$\lambda$, mode C1 joins a new family of spiral modes, which are
accumulated near the galactic center (see
\S\ref{sec:variations-of-lambda}). 

Reducing $\alpha$ increases the abundance of dark matter and
according to \cite{T81} and JH the growth rate of modes should decrease.
My calculations show that by reducing $\alpha$, spiral modes are
affected sooner and more effective than the bar mode, and they join
to the stationary modes, one by one from the location of the
bifurcation point until the whole S-family disappears. This is
a generic scenario for all $\lambda \ge 1$ models regardless of the
disk temperature controlled by $N$. Solid lines in
Figure \ref{pic:spectrum-m2-unstable}{\em b} show the {\it eigenfrequency
loci} of a model with $(N,\lambda)=(6,1)$ as $\alpha$ increases from
$0.2$ to $0.42$. It is evident that mode B1 is destabilized through
a pitchfork bifurcation while the loci of modes B2 and C1, and S-modes
exhibit a tangent bifurcation. All modes except mode G are stable
for $\alpha <0.23$. Surprisingly, mode G resists against stabilization
even for very small values of $\alpha$. This indicates that 
mode G is not characterized by the fraction of dark to luminous 
matter. In \S\ref{sec:variations-of-lambda} and \S\ref{sec:variation-of-N}, 
I will show that this mode is highly sensitive to the variations of 
$\lambda$ and $N$. According to my computations 
(e.g., Figure \ref{pic:spectrum-m2-unstable}{\em b}), by increasing 
$\alpha$ all S-modes are born at the same bifurcation frequency 
$\omega_{S}\approx 0.43$, but mode B2 comes out from a van Kampen 
mode with $\omega_{B}\approx 0.83$. This result completely rules out
any skepticism that mode B2 is a member of S-family.  

\begin{figure*}
\plotone{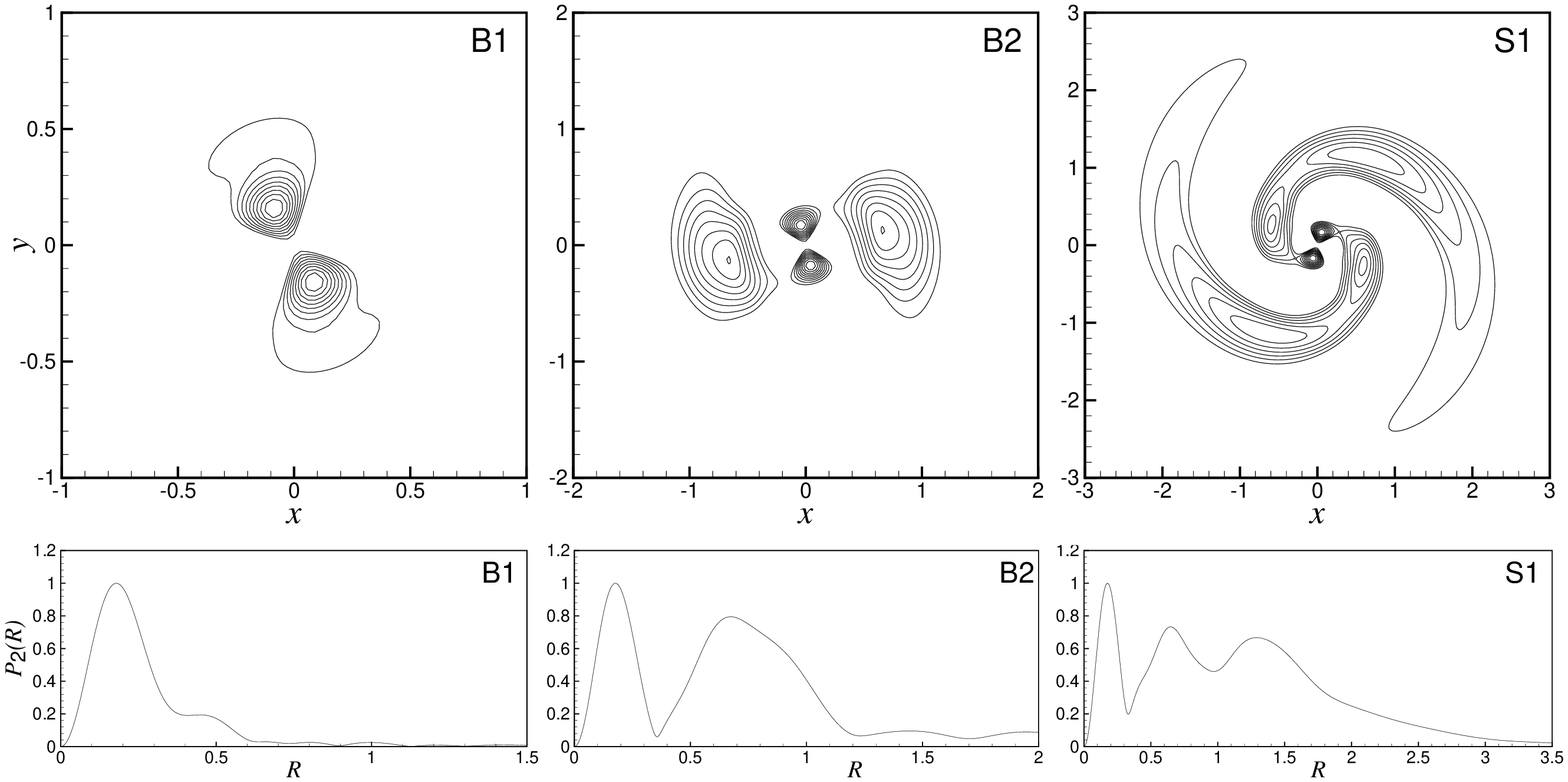}
\plotone{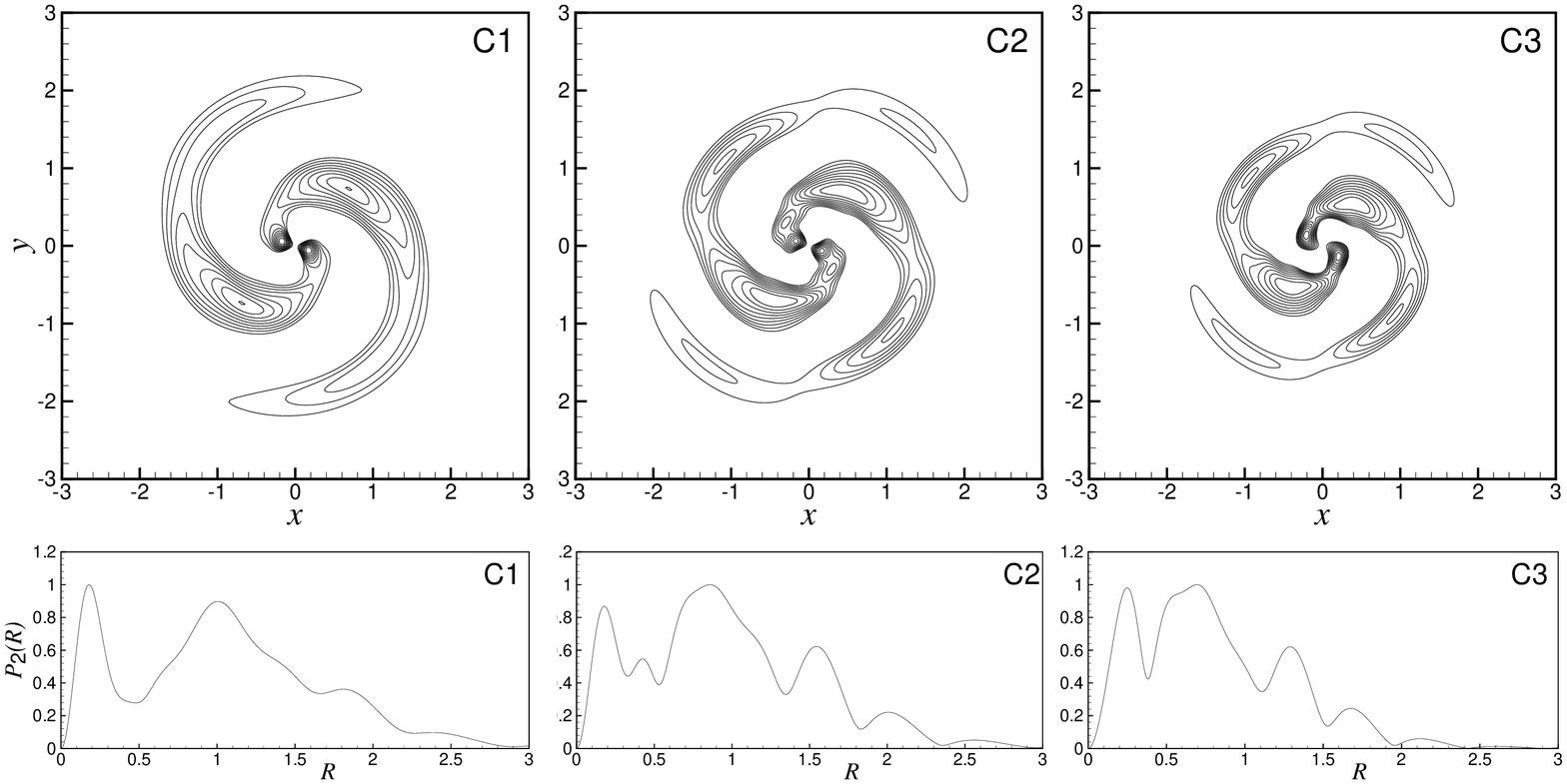}
\caption{Same as Figure \ref{pic:mode-shape-m2} but for a model
with $(\lambda,\alpha)=(0.625,0.34)$. The fastest growing spiral
mode, S1, has also been found by JH. Modes C1, C2 and C3 have
emerged due to dark matter presence at the galactic center.
They shrink to central regions and grow slower as their pattern
speed increases.
\label{pic:mode-shape-m2-SR34}}
\end{figure*}

Figure \ref{pic:mode-shape-m2} displays the wave patterns and
amplitude functions of modes B1, B2, G, S1, S2 and S3. It is seen
that the patterns of S-modes rotate slower and become more extensive 
as the mode number increases. Mode S6, which is at the bifurcation 
point of the spiral family, has the largest extent. Eight wave packets 
of this mode are distributed by a phase shift of 90 degrees along 
major spiral arms. Mode G has at most two density peaks on its major 
spiral arms but the magnitude of its second peak increases as the disk 
is cooled. Modes B1, B2, S1, S2 and S3 are, respectively, analogous to 
modes A, B, C, E and F of a Gaussian disk explored in \citet{T81}. 
There are three low-speed modes in Figure 11 of \citet{T81} that 
have not been labeled, but they are analogs of modes S4, S5 and S6. 
Mode G and Toomre's mode D also have some similarities but are of 
different origins (see \S\ref{sec:final-discussions}). None of them
can be stabilized only by increasing the fraction of dark matter. 

Figure \ref{pic:spectrum-m2-unstable}{\em a} shows that the 
fundamental mode obtained by JH coincides 
with mode B1. The wave pattern of mode B1 (displayed in Figure
\ref{pic:mode-shape-m2}) is identical to the mode shape computed
using Kalnajs's theory and demonstrated in Figure 8 of JH. JH found a
secondary mode which lies between modes B2 and G. That mode is also
a double-peaked spiral and it is not easy to identify its true
nature unless we investigate its evolution as the model parameters 
vary. By comparing Figure 10{\em a} of JH with 
Figure \ref{pic:spectrum-m2-unstable}{\em b}, one can see that both 
mode B2 and the secondary mode of JH are destabilized through a 
tangent bifurcation while mode G has a different nature. 
The bifurcation frequency of mode B2 that I find  
($\omega_{B}\approx 0.83$) matches very well with the frequency 
of the stabilized secondary mode of JH (see Figure 10{\em a} in JH
but note that their vertical axis indicates $\Omega_p=\omega_R/2$). 
Therefore, I conclude that the secondary mode of JH is indeed 
mode B2 although it seems to be closer to mode G. The existing 
discrepancy is due to different length scale of Clutton-Brock 
functions that JH have used for finding the secondary mode. 
By adjusting $b$ one can improve the location of B2. However, 
this is an unnecessary attempt given the fact that mode B2 has 
already been identified, and the computation accuracy of other 
eigenfrequencies has an impressive level.  

\subsection{Variations of
$\lambda$} \label{sec:variations-of-lambda}

The parameter $\lambda$ controls the density profile of the dark
matter component, specifically near the galactic center. The
fraction of dark to luminous matter has its minimum value in
marginal models with $\alpha\approx \alpha_{cr}$. I choose a
marginal $\lambda <1$ model with
$(N,\lambda,\alpha)=(6,0.625,0.34)$, which has also been investigated 
by JH. Figure \ref{pic:spectrum-m2-vary-lambda}{\em a} shows the portion 
of the spectrum that contains complex eigenfrequencies of this model. 
The spectrum has been computed for $b=1.5$. 
Although $m=2$ bar and spiral modes survive 
in this model, their pattern speeds and growth rates drop considerably.
Mode G has been wiped out of existence by dark matter penetration into 
the center, and four unstable modes (C1, C2, C3 and C4) have emerged 
that constitute a new family of spiral modes. They populate the central 
regions of the disk in most of $\lambda <1$ models. Again, the location 
of eigenfrequencies obtained by JH have been marked by circles. 
The agreement between the results of JH and the present study is very 
good and the variance is less than $2\%$.

The population of spiral modes is changed by varying $\alpha$, 
and the eigenfrequencies of unstable modes are altered 
significantly. Solid lines in Figure
\ref{pic:spectrum-m2-vary-lambda}{\em a} show the loci of growing
modes as $\alpha$ increases from $0.2$ to $\alpha_{cr}\approx 0.36$.
Similar to the previous $\lambda=1$ model, S-modes and mode B1 are
destabilized through tangent and pitchfork bifurcations,
respectively. All C-modes are born by a pitchfork bifurcation 
although some minor modes of the same nature come and go as $\alpha$
varies. The loci of modes B1 and S1 (in Figure 
\ref{pic:spectrum-m2-vary-lambda}{\em a}) are in harmony with the
results of Kalnajs's method plotted in Figure 10{\em b} of JH. 
It is noted that the locus of mode B1 steeply joins the real axis,
well before stabilizing the S-modes. This is how slowly growing 
spirals may dominate a stellar disk.   

There are no new families of growing modes in $\lambda>1$ models.
Dark matter in these models induces a rising rotation curve on the
disk stars (see Figure 6 in JH) and the population of S-modes declines. 
The growth rate of mode B2 increases proportional to $\lambda$, but 
that of mode G falls off although mode G is still robust against 
the variations of $\alpha$. Mode G has its maximum growth rate in
$\lambda=1$ models, which suggests that it must be a self-gravitating 
response of the luminous matter that involves only the potential 
of the disk, $V_D$. The function $P_2(R)$ of mode C1 loses 
its minor peaks and becomes smoother as $\lambda$ increases.  
Figure \ref{pic:spectrum-m2-vary-lambda}{\em b} shows the  
eigenfrequency loci of models with $(N,\lambda)=(6,2)$ as 
$\alpha$ increases from $0.25$ to $\alpha_{cr}\approx 0.6$. 
The eigenfrequency loci of modes C1 and B2, and S-modes (as $\alpha$ 
varies) are similar to $\lambda=1$ models, but the locus of mode B1 
loses its steepness and stretches towards small pattern speeds in 
an approximately linear form until it joins the real $\omega_R$-axis.
The bifurcation frequency of mode B2 differs from S-modes and it is 
larger. 
 
Figure \ref{pic:mode-shape-m2-SR34} shows the wave patterns of modes
B1, B2, S1, C1, C2 and C3 for the model with 
$(N,\lambda,\alpha)=(6,0.625,0.34)$. The (isolated) mode B1 is still 
a single-peaked bar although its edge is more extensive as the flat
part of its $P_2(R)$ plot indicates. Mode S1 is a triple-peaked
spiral (as before) and mode B2 is being stabilized ($\omega_{\rm
B2}=0.775+0.007{\rm i}$). It is seen that mode B2 has a bar-like
structure, which justifies its classification as the secondary bar
mode. The pattern of S1 and its $P_2(R)$ plot can be compared with
Figure 9 in JH. The agreement is quite satisfactory. There is a
remarkable difference between the patterns of C- and S-modes
although both families have spiral structures. In contrast to
S-modes that become more extensive as their growth rate decays,
C-modes are shrunk to central regions because their pattern speed
increases. C-modes are also a bifurcating family, but their
bifurcation point lies on large pattern speeds associated with the
azimuthal frequency ($\Omega_\phi$) of central stars.  
 
The parameters $\alpha$ and $\lambda$ are essentially controlling 
the fraction and density profile of dark matter component, respectively. 
However, the radial velocity dispersion $\sigma_R$ of the equilibrium
state is also playing an important role in the perturbed dynamics. 
$\sigma_R$ is an indicator of the initial temperature of the disk. 
\citet{ER98b} had already pointed out that the pitch angle of spiral 
patterns decreases when $\sigma_R\rightarrow 0$ (see their Figure 7). 
Apart from this morphological implication, can the variation in the 
disk temperature affect the modal content? In the following subsection, 
I trace the evolution of growing waves by changing the disk temperature 
and show that the population of S-modes is larger in rotationally 
supported, cold disks.

\begin{figure*}
\plottwo{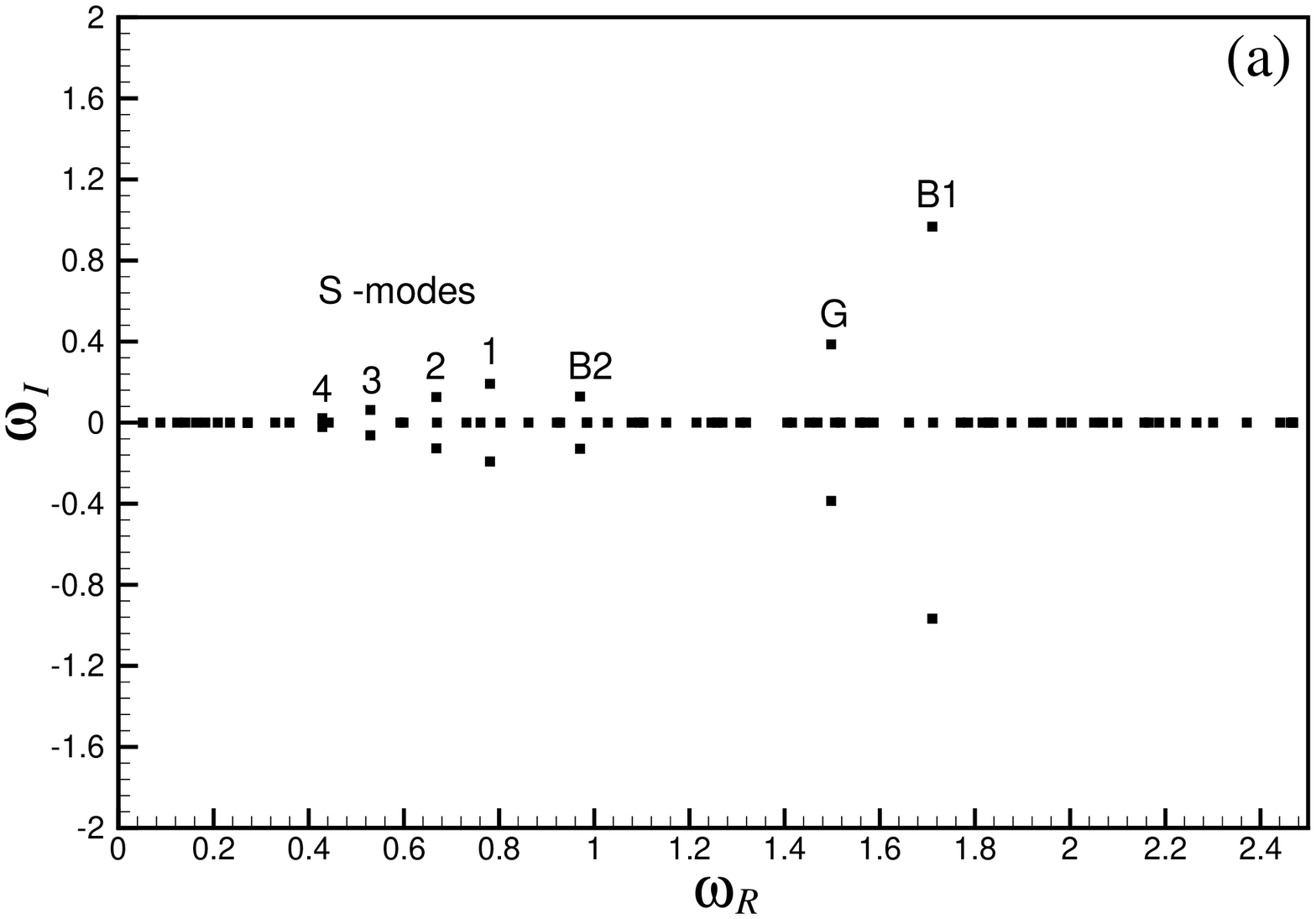}{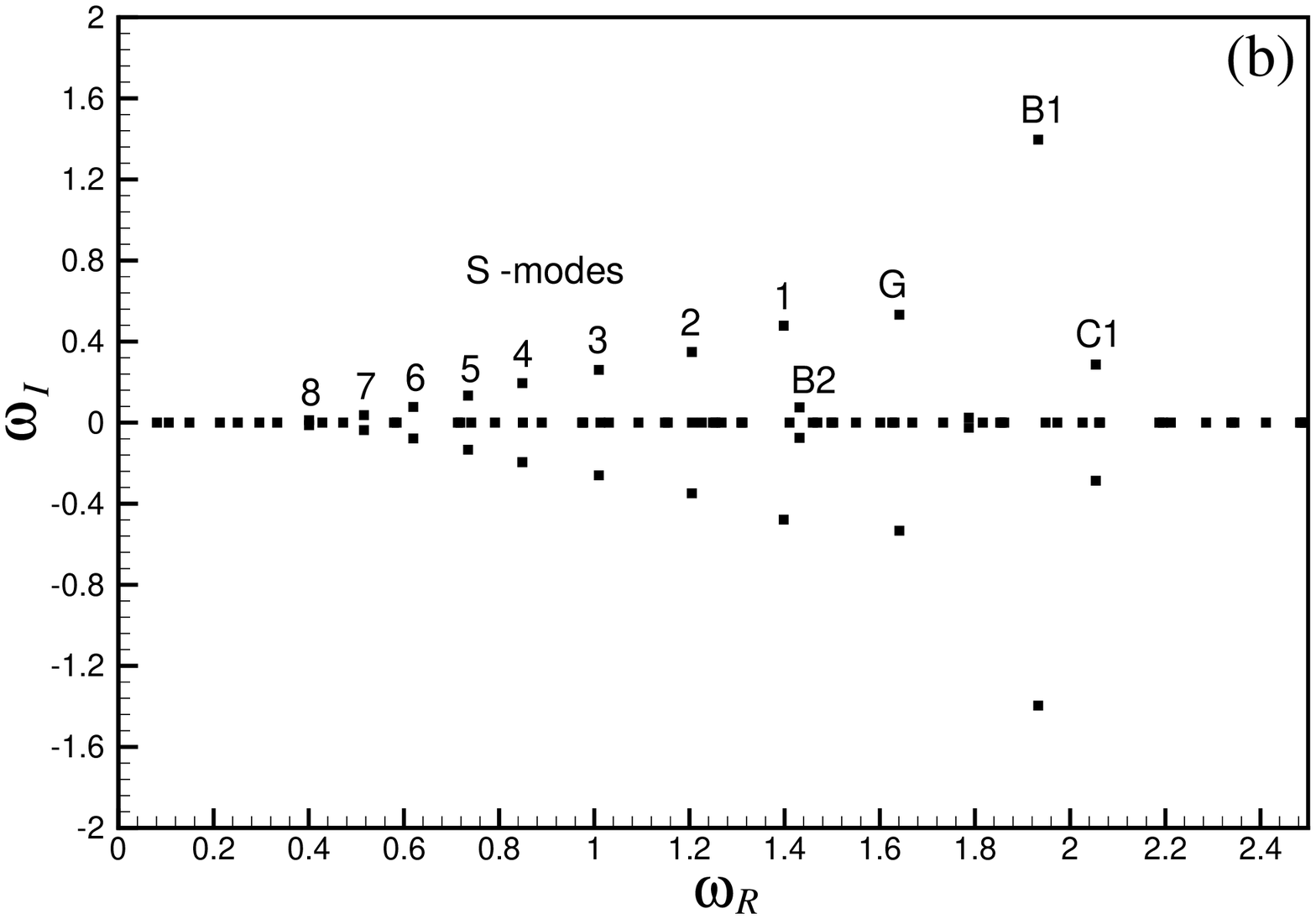}
 \plottwo{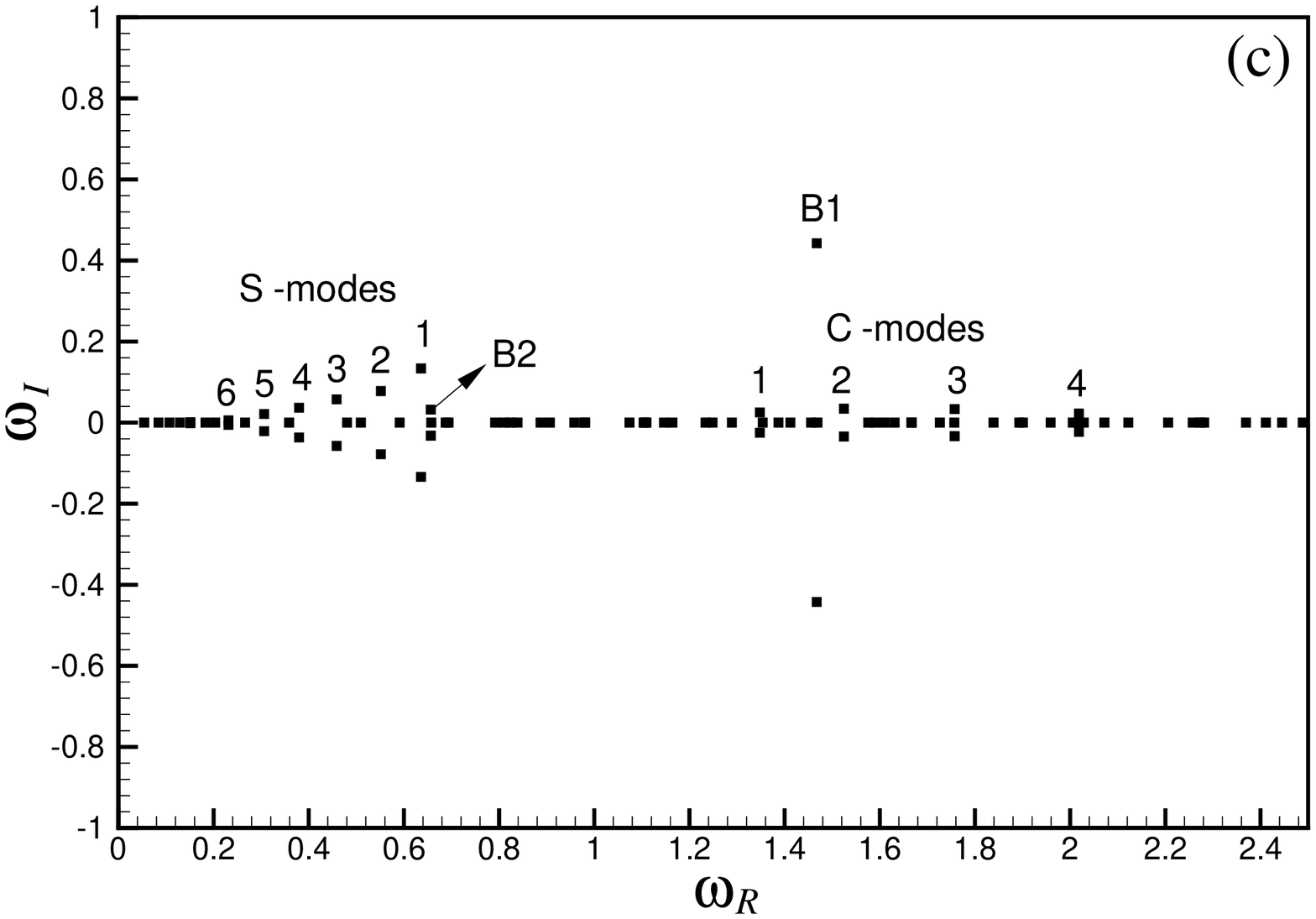}{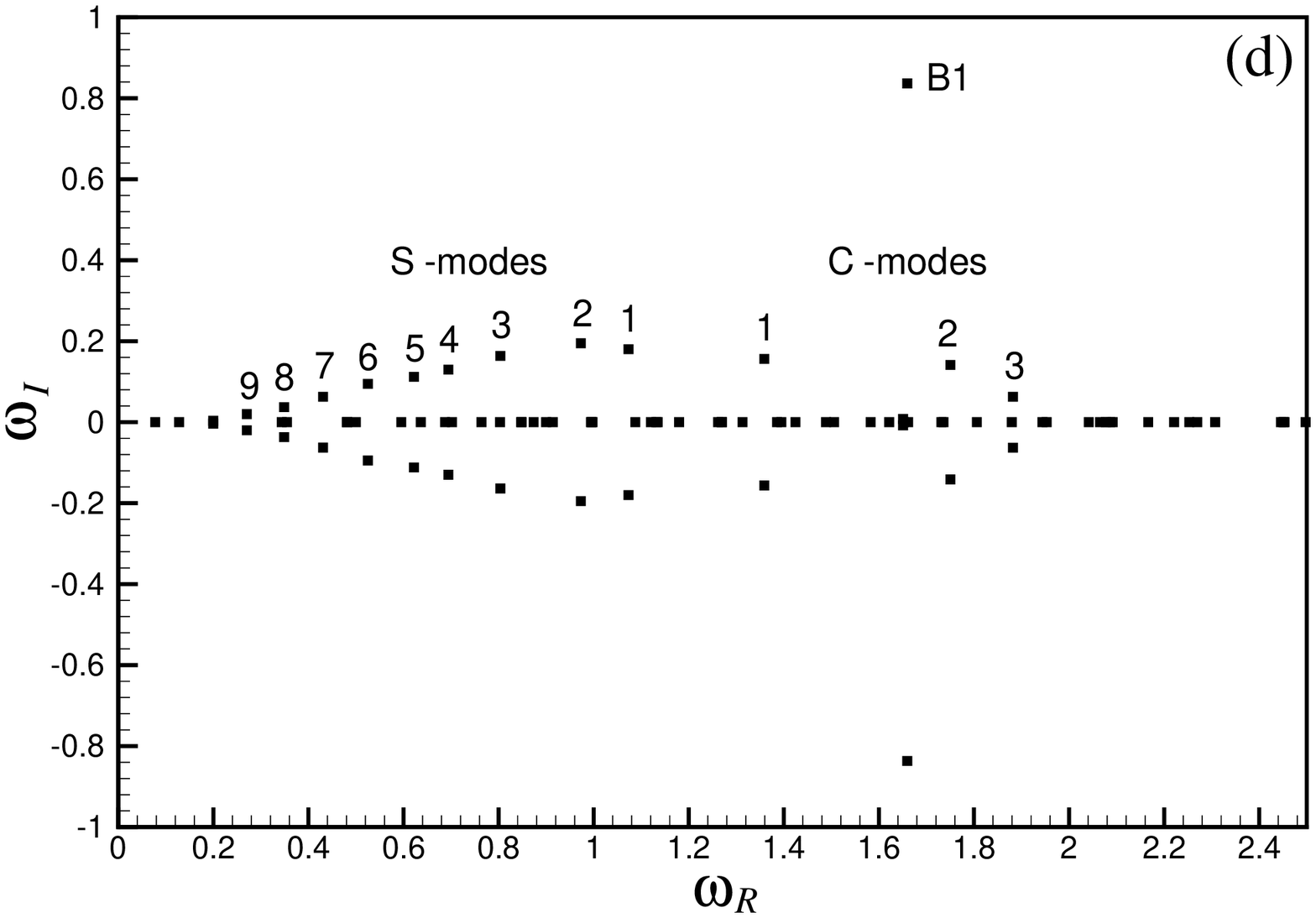}
 \caption{The evolution of the eigenfrequency spectra as the
disk temperature drops from left ($N=4$) to right ($N=8$) panels.
Top and bottom panels correspond to $(\lambda,\alpha)=(1,0.42)$
and $(\lambda,\alpha)=(0.625,0.34)$, respectively.
\label{pic:spectrum-vary-N}}
\end{figure*}
\begin{figure}
\plotone{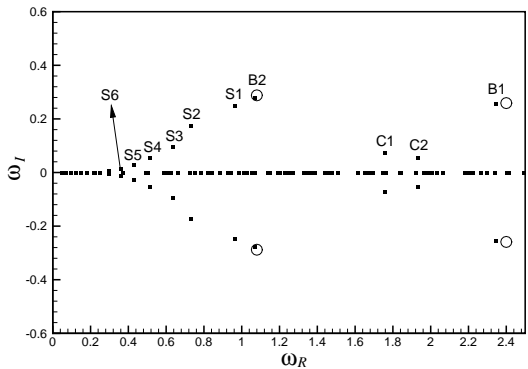}
 \caption{Eigenfrequency spectrum
of a cutout model with $L_0=0.1$ and $(N,\lambda,\alpha)=(6,1,0.42)$. 
Circles show the eigenfrequencies found by JH using Kalnajs's theory. 
\label{pic:spectrum-cutout}}
\end{figure} 

\subsection{Variations of the Disk Temperature}
 \label{sec:variation-of-N}

The parameter $N$ of the DFs of JH controls the disk temperature by
adjusting the size of the isotropic core and the population of near
circular orbits. As $N$ increases, the streaming velocity $\langle
v_{\phi}\rangle$ approaches the rotational velocity of circular
orbits and the stellar disk is cooled. Figure \ref{pic:spectrum-vary-N}
displays the eigenfrequencies of previous $(\lambda,\alpha)=(1,0.42)$
and $(\lambda,\alpha)=(0.625,0.34)$ models for $N=4$ and $N=8$. The
spectra for the intermediate value of $N=6$ have already been shown
in Figures \ref{pic:spectrum-m2-unstable}{\em a} and
\ref{pic:spectrum-m2-vary-lambda}{\em a}.

Increasing $N$ gives birth to more S-modes while the bifurcation
point of the family is preserved. As a new member is born at the
bifurcation point, other members including mode S1, are pushed away
from the real axis on a curved path. This behavior is observed in
both models but the branch of S-family in the model with
$(\lambda,\alpha)=(0.625,0.34)$ stays closer to the real axis than
the other model. The growth rates of C-modes increase remarkably as
the disk is cooled. Despite mode B1 which rotates and grows faster
in cold disks, mode B2 grows faster in warmer disks. Variation
in the disk temperature changes the eigenfrequency of mode G more
effective than what $\alpha$ could, but nothing is more influential
than the role of $\lambda$.  

Another consequence of cooling the stellar disk is that $m=0$ waves
are no longer stable. The parameter $Q$ of Toomre was marginally
greater than unity for $N=6$ models. For $N=8$, I find $Q<1$
over an annular region because the plot of $Q$ versus $R$ exhibits a
minimum at some finite radius (e.g., Figure 7 of JH). For instance, 
I find three growing
$m=0$ modes for the model $(N,\lambda,\alpha)=(8,1,0.42)$. They
correspond to pure complex eigenfrequencies $\omega_1=0.621{\rm i}$,
$\omega_2=0.494{\rm i}$ and $\omega_3=0.238{\rm i}$. Mode shapes
have (obviously) ringed structures but the number of rings, which is
identical to the number of peaks of $P_0(R)$, depends on the growth
rate. The modes associated with $\omega_1$, $\omega_2$ and
$\omega_3$ have three, four and five rings, respectively. Ring modes
are very sensitive to the variations of model parameters and they
are suppressed by decreasing $\lambda$ and $\alpha$.

\subsection{The Effect of an Inner Cutout}
\label{sec:cutout-models}

In order to simulate an immobile bulge, which does not respond to
density perturbations, JH utilized an inner cutout function of the
form
\begin{equation}
H_{\rm cut}=1-e^{-\left(J_\phi/L_0\right)^2},
\end{equation}
where $L_0$ is an angular momentum scale. Multiplying $H_{\rm cut}$
by the self-consistent DF of the equilibrium state, prohibits the
stars with $J_{\phi}< L_0$ from participating in the perturbed
dynamics. Consequently, incoming waves are reflected at some finite
radius and the innermost wave packets of multiple-peaked modes are
diminished. My calculations show that all S-modes survive in cutout 
models, mode G disappears, and the growth rate of mode B2 increases.
The pattern speed of mode B1 is boosted so that the corotation
resonance is destroyed, but its growth rate drops drastically.
Figure \ref{pic:spectrum-cutout} shows the eigenfrequency spectrum
of a model with $L_0=0.1$ and $(N,\lambda,\alpha)=(6,1,0.42)$. 
Circles show the eigenfrequencies found by JH. Again, the agreement
between the results of JH and the present work is very good. The
reason that I have identified mode B2 as the second member of
B-family, and not the most unstable S-mode, is that its $P_2(R)$
function has an evolved double-peaked structure (see Figure 11 in JH)
and its locus versus $\alpha$ does not emerge from the same bifurcation 
frequency of S-modes. Disappearance of mode G in cutout models confirms 
my earlier note that it is a self-gravitating mode.

\section{DISCUSSIONS}
\label{sec:final-discussions}

There are similarities between mode G of this study and Toomre's (1981) 
mode D. Both of these modes resist against stabilization by increasing 
the fraction of dark to luminous matter and they have at most double 
peaks on their spiral arms. Nonetheless, these modes are not the same 
because mode G is amplified through a feedback from the galactic center 
but Toomre's mode D has been identified as an edge mode. A question 
remains to be answered: why Toomre (1981) did not detect mode G and I 
do not find an edge mode? The most convincing explanation is that to 
excite a self-gravitating wave inside the core of the stellar component, 
the governing potential in that region should mainly come from the 
self-gravity of stars. This requirement is fulfilled in my $\lambda\ge 1$ 
models for $R<R_D$. However, the completely flat rotation curve imposed 
by Toomre (1981) nowhere follows the rotational velocity induced by the 
self-gravity of stars and it prohibits the Gaussian disk from developing 
a G-like mode. On the other hand, I don't find an edge mode because the 
density profile of the cored exponential disk does not decay as steep as 
the Gaussian disk to create an outer boundary at some finite radius for 
reflecting the outgoing waves.    

Similar to the first order analysis of \S\ref{sec::linear-theory},
Polyachenko's (2005) approach results in the full spectrum of
eigenfrequencies for a given azimuthal wavenumber. There are some
differences between his method and the present formulation.
Polyachenko directly uses Poisson's integral to establish a
point-wise relation in the action space between the Fourier
components of the perturbed DF and its self-consistent
potential. Combination of equations (5) and (9)
in his paper is analogous to equation (\ref{eq:expansion-V1-vs-dmlj})
in this paper. The main departure of the two theories is in the way
that the linearized CBE is treated. Polyachenko forces a point-wise
fulfillment of the CBE in the action space while the present method
works with a weighted residual form of the CBE.

A point-wise formulation poses a challenge for the numerical
calculation of the eigenvalues and their conjugate eigenvectors.
According to the bar charts of JH, at least ten Fourier
components ($-3 \le l \le 6$) are needed in the $\theta_R$-direction
to assure a credible convergence of $f_1$ in a typical soft-centered
galaxy model. Therefore, if one chooses a grid of $n_a\times n_a$ in the
action space, Polyachenko's eigenvector $\textbf{\textit{F}}$ will have a
dimension of $10\times n_a\times n_a$. Therefore, for a {\it very coarse}
grid with $n_a=21$ that Polyachenko uses, the unknown eigenvector
will have a dimension of 4410. This number must be compared with
the dimension of $\textbf{\textit{z}}_0$ in equation
(\ref{eq:eigenvalue-problem-two}). That is indeed $n_{\rm max}=336$
for the most accurate calculations carried out by setting
$(l_{\rm max},j_{\rm max})=(10,15)$ which means that $21$
Fourier components in the $\theta_R$-direction and $16$
expansion terms in the $R$-direction have been taken into account.
Noting that the definite integrals ${\cal I}^{ml}_{jk}$ and
$\Lambda^{ml}_{jk}$ are independently evaluated over the action
space with any desired accuracy, the present theory proves to
be more efficient for eigenmode calculation (in the linear regime)
than other existing alternatives.

The agreement between the results of this work and those of JH, who
have used Kalnajs's method, is impressive. There is only a
discrepancy in the results for a double-peaked spiral mode of
$\lambda=1$ models. In fact, these models have two double-peaked
modes, modes B2 and G, and JH find mode B2. The origin of
discrepancies was attributed to the length scale of Clutton-Brock
functions, $b$, which is a fixed number for the whole spectrum of 
a given azimuthal wavenumber. Provided that JH optimized $b$ for each
growing mode that they calculated (see also \S\ref{sec:maximal-disk}), 
some minor deviations from the results of this paper are reasonable. 
In most cases the algorithm used by JH converges to mode B1 and the 
fastest rotating S-mode. They capture mode B2 only if
its growth rate is large enough. Other modes remain unexplored
because Newton's method needs an initial guess of $\omega$,
which has a little chance to be in the basin of attraction of the
other members of S-family. The separation of eigenfrequencies near
the bifurcation point of S-modes is very small and one could
anticipate complex boundaries for the basins of attraction of these
eigenfrequencies. Thus, there is no guarantee that successive
Newton's iterations keep an estimated eigenfrequency on the same
basin that it was initially. Nevertheless, in Kalnajs's formulation,
a systematic search for all growing modes is possible by introducing
the mathematical eigenvalue \citep{Z76,ER98b} and investigating its
loci as the pattern speed and growth rate vary.

\section{CONCLUSIONS}
\label{sec:conclusions}

After three decades of Kalnajs's (1977) publication, it was not
known exactly whether growing modes of stellar systems appear as
distinct roots in the eigenfrequency space or they belong to
continuous families as van Kampen modes do. In this paper, I
attempted to answer this question using the Galerkin projection
of the CBE and unveiled the full eigenfrequency spectrum of a
stellar disk. I showed that similar to gaseous disks \citep{AJ06},
majority of growing modes emerge as {\it discrete families} 
through a bifurcation from stationary modes. There are some 
exceptions for this rule, the most important of which are 
the isolated bar and G modes.  

The model that I used to test my method allows for dark matter
presence as a spherical component, whose potential inside the
galactic disk contributes to the rotational velocity of stars.
By varying the parameters of the model, and investigating the
eigenfrequencies and their associated mode shapes, I showed that
it is not the fraction of dark to luminous matter that controls
the variety of growing modes. What determines that variety is
indeed the shape of the dark matter density profile controlled
by the parameter $\lambda=R_C/R_D$. My survey in the parameter
space revealed that the concentration of dark matter in the 
galactic center ($\lambda <1$) destroys mode G and weakens the 
growth of B-modes substantially. Emergence of spiral C-modes 
that accumulate near the galactic center is another remarkable 
consequence of dark matter presence in central regions of a 
cored stellar disk.  

Although the solution of the Galerkin system showed a credible
convergence of the series expansions, the existence of strong
solutions for the CBE, in its full nonlinear form, is still an
open problem. It has been known for years that van Kampen modes
make a complete set \citep{C59}, and therefore, they may be used
for a series representation of stationary oscillations. But there is
not a mathematical proof for the completeness of the discrete families
of growing modes. In other words, whether an observed galaxy can be
assembled using the modes of a linear eigensystem, requires further
analysis.

In the second part of this study, I will investigate the mechanisms
of wave interactions in the nonlinear regime and will probe the mass
and angular momentum transfer between waves of different Fourier
numbers.

\acknowledgments I am indebted to Chris Hunter for his instructive
and valuable comments since the beginning of this work. I also thank
the referee for helpful suggestions that improved the presentation
of the results. This work was partially supported by the Research 
Vice-Presidency at Sharif University of Technology.

\appendix

\section{WEIGHTED RESIDUAL FORM OF THE COLLISIONLESS BOLTZMANN EQUATION} 
\label{app::Petrov-Galerkin}

Let me define a nonlinear operator ${\cal A}$ and denote 
$\textbf{\textit{u}}^{(\ell)}$ as the $\ell$th prolongation \citep{OL93}
of the physical quantity $u$ in the domain of independent 
variables. Assume a (nonlinear) partial differential equation
\begin{equation}
{\cal A} \left ( u^{(\ell)}, x,t \right )=0, 
\label{eq:general-operator-prolonged}
\end{equation}
and its associated initial and boundary conditions that govern 
the evolution of $u(x,t)$ in the domain of the 
spatial variable $x$ and the time $t$. A weighted
residual method \citep{F72} attempts to find an approximate solution 
of the form 
\begin{equation}
u(x,t)=
\sum_{k=1}^{k_{\rm max}} a_{k}(t) 
\varphi_{k}(x),
\label{eq:expansion-for-u-general}
\end{equation}
through determining the time-dependent functions $a_k(t)$ for a given 
set of trial (basis) functions $\varphi_{k}(x)$.
The trial functions should satisfy the boundary conditions and be 
linearly independent. Using (\ref{eq:expansion-for-u-general}) and taking 
the inner product of (\ref{eq:general-operator-prolonged}) by some weighting 
functions $W_{k'}(x)$, yield the determining equations 
of $a_k(t)$ as
\begin{equation}
\left ( {\cal A},W_{k'} \right ) \equiv 
\int {\cal A} W_{k'} dx=0,~~
k'=1,2,\cdots,k_{\rm max}. 
\label{eq:weighted-residual-form-general}
\end{equation}
There are several procedures for choosing $W_{k'}(x)$
and each procedure has its own name. The method with $W_{k'}=\varphi_{k'}$ 
is called the Bubnov-Galerkin, or simply the Galerkin method. The 
Petrov-Galerkin method is associated with $W_{k'}\not=\varphi_{k'}$.
The well-known collocation method uses Dirac's delta functions for 
the weighting purpose. There is an alternative interpretation for the 
inner product $\left ({\cal A},W_{k'} \right )=0$. That is projecting 
the equation ${\cal A}=0$ on a subspace spanned by the weighting function 
$W_{k'}$. Therefore, equation (\ref{eq:weighted-residual-form-general}) 
is often called the {\it Galerkin projection} of 
(\ref{eq:general-operator-prolonged}). In what follows, I use the 
Petrov-Galerkin method and construct the weighted residual form of 
the CBE.   
   
Assume the functions $U(\Theta,\textbf{\textit{J}})$ and
$V(\Theta,\textbf{\textit{J}})$, and define their inner product over the
action-angle space as
\begin{equation}
\left (U,V \right )=\int\int
U(\Theta,\textbf{\textit{J}})V(\Theta,\textbf{\textit{J}})
 d \textbf{\textit{J}} d \Theta.
\end{equation}
Taking the inner product of the perturbed CBE by the weighting functions
$W^{ml}_{j}(\Theta,\textbf{\textit{J}})=
\Psi^{ml}_{j}(\textbf{\textit{J}})e^{-{\rm i}
\left (l\theta_R+m\theta_\phi \right)}$ gives
\begin{equation}
\left (\frac{\partial f_1}{\partial t},W^{ml}_{j}\right )=
-\left ( \left [f_1,{\cal H}_0 \right ],W^{ml}_{j}\right )
-\left ( \left [f_0,{\cal H}_1 \right ],W^{ml}_{j} \right )-
\left ( \left [f_1,{\cal H}_1 \right ],W^{ml}_{j}\right ).
\label{eq:CBE-weighted-residual-1}
\end{equation}
Note that the CBE is the governing equation of the perturbed DF whose
trial functions are $\Phi^{ml}_j(\textbf{\textit{J}})$. With my choice 
of the weighting function (as above) I am following the 
Petrov-Galerkin method. On substituting (\ref{eq:expansion-f1}) and
(\ref{eq:expansion-V1-vs-dmlj}) in
(\ref{eq:CBE-weighted-residual-1}) and after some rearrangements of
summations, one obtains  
\begin{eqnarray}
&{}& {\rm i}\sum_{m',l'}^{}\sum_{j'}^{} \delta_{m,m'}\delta_{l,l'}{{\rm
d}\over  d t}d^{m'l'}_{j'}(t)\int  d \textbf{\textit{J}}
\Psi^{ml}_{j}(\textbf{\textit{J}})\Phi^{m'l'}_{j'}(\textbf{\textit{J}})=
\nonumber \\
&{}& \sum_{m',l'}^{}\sum_{j'}^{}
\delta_{m,m'}\delta_{l,l'}d^{m'l'}_{j'}(t)
\int  d \textbf{\textit{J}} \left ( l'\Omega_R+m'\Omega_\phi \right
)\Psi^{ml}_{j}(\textbf{\textit{J}})\Phi^{m'l'}_{j'}(\textbf{\textit{J}})
\nonumber \\
&-& \sum_{m',l'}^{}\sum_{j'}^{} \delta_{m,m'} d^{m'l'}_{j'}(t)
\sum_{k}^{} \left [ {4\pi^2\over D_k(m')}\right ] \Lambda^{m'l'}_{kj'}
\int  d \textbf{\textit{J}} \left ( l{\partial f_0\over \partial J_R}
+m'{\partial f_0\over \partial J_{\phi}} \right
)\Psi^{m'l}_{j}(\textbf{\textit{J}})\Psi^{m'l}_k(\textbf{\textit{J}})
\nonumber \\
&+& \sum_{m',l'}^{}\sum_{j'}^{}
    \sum_{m'',l''}^{}\sum_{j''}^{}
\delta_{m'',(m-m')} d^{m'l'}_{j'}(t) d^{m''l''}_{j''}(t)
\sum_{k}^{} \left [ {4\pi^2\over D_k(m'')} \right ]
\Lambda ^{m''l''}_{kj''} \nonumber \\
&\times& \Biggl [ \int  d \textbf{\textit{J}}
\Psi^{ml}_{j}(\textbf{\textit{J}})\Phi^{m'l'}_{j'}(\textbf{\textit{J}})
\left ( l'{\partial \over \partial J_R}+
        m'{\partial \over \partial J_\phi} \right )
\Psi^{m''(l-l')}_{k}(\textbf{\textit{J}}) \nonumber \\
&{}& \qquad -\int  d \textbf{\textit{J}}
\Psi^{ml}_{j}(\textbf{\textit{J}})
\Psi^{m''(l-l')}_{k}(\textbf{\textit{J}})
\left ( (l-l'){\partial \over \partial J_R} +
        m''{\partial \over \partial J_\phi} \right )
\Phi^{m'l'}_{j'}(\textbf{\textit{J}}) \Biggr ],
\label{eq:weighted-residual-expan}
\end{eqnarray}
where $-m_{\rm max}\le m,m',m''<m_{\rm max}$,
$-l_{\rm max}\le l,l',l''\le l_{\rm max}$ and $0 \le
j,j',j'',k\le j_{\rm max}$. Using equation (\ref{eq:map-mlj-to-i})
and carrying out the index mappings $(m,l,j)\rightarrow p$,
$(m',l',j')\rightarrow q$ and $(m'',l'',j'')\rightarrow r$ one may
introduce the arrays
\begin{eqnarray}
M_{pq}&=& \delta_{m,m'}\delta_{l,l'}\int  d \textbf{\textit{J}}
\Psi^{ml}_{j}(\textbf{\textit{J}})\Phi^{m'l'}_{j'}(\textbf{\textit{J}}),
\label{eq:M-tensor} \\
C_{pq}&=& \delta_{m,m'} \Biggl [ \delta_{l,l'}
\int  d \textbf{\textit{J}}
\left ( l'\Omega_R+m'\Omega_\phi \right )\Psi^{ml}_{j}(\textbf{\textit{J}})
\Phi^{m'l'}_{j'}(\textbf{\textit{J}}) \nonumber \\
&{}& \qquad \qquad -\sum_{k=0}^{j_{\rm max}}
\left [ {4\pi^2\over D_k(m')}\right ]
\Lambda^{m'l'}_{kj'} \int  d \textbf{\textit{J}} \left ( l{\partial
f_0\over \partial J_R} +m'{\partial f_0\over \partial J_{\phi}}
\right )\Psi^{m'l}_{j}(\textbf{\textit{J}})
\Psi^{m'l}_k(\textbf{\textit{J}})\Biggr ],
\label{eq:A-tensor} \\
K_{pqr}&=& \delta_{m',(m-m'')} \delta_{l',(l-l'')}
\sum_{k=0}^{j_{\rm max}} \left [ {4\pi^2\over
D_k(m'')} \right ] \Lambda ^{m''l''}_{kj''} \nonumber \\
&{}& \times \Biggl [ \int  d \textbf{\textit{J}}
\Psi^{ml}_{j}(\textbf{\textit{J}})
\Phi^{m'l'}_{j'}(\textbf{\textit{J}})
\left ( l'{\partial \over \partial J_R}+
        m'{\partial \over \partial J_\phi} \right )
\Psi^{m''l''}_{k}(\textbf{\textit{J}}) \nonumber \\
&{}& \qquad -\int  d \textbf{\textit{J}}
\Psi^{ml}_{j}(\textbf{\textit{J}})
\Psi^{m''l''}_{k}(\textbf{\textit{J}})
\left ( l''{\partial \over \partial J_R} +
        m''{\partial \over \partial J_\phi} \right )
\Phi^{m'l'}_{j'}(\textbf{\textit{J}}) \Biggr ].\label{eq:B-tensor}
\end{eqnarray}
Consequently, equation (\ref{eq:weighted-residual-expan}) takes the
following matrix form
\begin{equation}
{\rm i}\sum_{q=1}^{n_{\rm max}} M_{pq} { d \over  d t}z_q(t) =
\sum_{q=1}^{n_{\rm max}} C_{pq}z_q(t)+\sum_{q,r=1}^{n_{\rm max}}
K_{pqr}z_q(t)z_r(t),~~z_p(t)\equiv d^{ml}_j(t),~~
p=1,2,\cdots,n_{\rm max}.
\label{eq:ODE-for-z-one}
\end{equation}
Let the matrix $\textbf{\textit{M}}^{-1}=[M^{-1}_{pq}]$ be the inverse of
$\textbf{\textit{M}}=[M_{pq}]$ and left-multiply (\ref{eq:ODE-for-z-one})
by $\textbf{\textit{M}}^{-1}$ to get
\begin{equation}
{\rm i} { d \over  d t}z_p(t) =
\sum_{q=1}^{n_{\rm max}} A_{pq}z_q(t)+\sum_{q,r=1}^{n_{\rm max}} B_{pqr}
z_q(t)z_r(t),~~\textbf{\textit{A}}=\textbf{\textit{M}}^{-1}
\cdot \textbf{\textit{C}},~~
B_{pqr}=\sum_{s=1}^{n_{\rm max}}M^{-1}_{ps}K_{sqr}.
\label{eq:ODE-for-z-two}
\end{equation}
Evaluation of the integrands in (\ref{eq:B-tensor}) will be
considerably simplified if one avoids the partial derivatives
of $\Phi^{ml}_j(\textbf{\textit{J}})$ through integrating (\ref{eq:B-tensor})
by parts. That gives
\begin{eqnarray}
K_{pqr}= &{}& \delta_{m',(m-m'')} \delta_{l',(l-l'')}
\sum_{k=0}^{j_{\rm max}} \left [ {4\pi^2\over
D_k(m'')} \right ] \Lambda ^{m''l''}_{kj''} \nonumber \\
\times \biggl \{ &{}& \int {\rm
d}\textbf{\textit{J}} \Psi^{ml}_{j}(\textbf{\textit{J}})
\Phi^{m'l'}_{j'}(\textbf{\textit{J}})
\left ( l{\partial \over \partial J_R}+
        m{\partial \over \partial J_\phi} \right )
\Psi^{m''l''}_{k}(\textbf{\textit{J}}) \nonumber \\
&+& \int  d \textbf{\textit{J}}
\Phi^{m'l'}_{j'}(\textbf{\textit{J}})
\Psi^{m''l''}_{k}(\textbf{\textit{J}})
\left ( l''{\partial \over \partial J_R} +
        m''{\partial \over \partial J_\phi} \right )
\Psi^{ml}_{j}(\textbf{\textit{J}})
\biggr \}.\label{eq:B-tensor-by-parts}
\end{eqnarray}
When all stars move on prograde orbits, the equilibrium DF takes the
form $f_0(\textbf{\textit{J}})=H(J_\phi)f^P_0(\textbf{\textit{J}})$
where $H$ is the
Heaviside function. Upon using (\ref{eq:trial-functions-two}),
this contributes a term including Dirac's delta function
$\delta(J_\phi)$ to the trial functions. Thus, the following
boundary terms
\begin{eqnarray}
K^b_{pqr} \! &=& \! \delta_{m',(m-m'')} \delta_{l',(l-l'')}
\sum_{k=0}^{j_{\rm max}}
\left [ {4\pi^2\over D_k(m'')} \right ]
\Lambda ^{m''l''}_{kj''} \nonumber \\
\times \Biggl \{ \int_{0}^{\infty} \!\! &  d J_R &
\left [ {m' f^P_0(\textbf{\textit{J}})\Psi^{ml}_{j}(\textbf{\textit{J}})
                          \Psi^{m'l'}_{j'}(\textbf{\textit{J}}) \over
         l'\Omega_R(\textbf{\textit{J}})+m'\Omega_\phi(\textbf{\textit{J}})
        }
\left ( l{\partial \over \partial J_R}+
        m{\partial \over \partial J_\phi} \right )
\Psi^{m''l''}_{k}(\textbf{\textit{J}}) \right ]_{J_{\phi}=0} \nonumber \\
+ \int_{0}^{\infty} \!\! &  d J_R & \left [
        {m' f^P_0(\textbf{\textit{J}}) \Psi^{m''l''}_{k}(\textbf{\textit{J}})
                             \Psi^{m'l'}_{j'}(\textbf{\textit{J}})
   \over l'\Omega_R(\textbf{\textit{J}})+m'\Omega_\phi(\textbf{\textit{J}})
        }
\left ( l''{\partial \over
\partial J_R} +
        m''{\partial \over \partial J_\phi} \right )
\Psi^{ml}_{j}(\textbf{\textit{J}}) \right ]_{J_\phi=0}
 \Biggr \}, \label{eq:B-tensor-boundary}
\end{eqnarray}
must be added to $K_{pqr}$ when the equilibrium
disk is unidirectional. The partial derivatives of
$\Psi^{ml}_j(\textbf{\textit{J}})$ needed for equations
(\ref{eq:B-tensor-by-parts}) and (\ref{eq:B-tensor-boundary})
are calculated by differentiating equation (\ref{eq:fourier-coeffs})
partially with respect to an action:
\begin{eqnarray}
{\partial \Psi^{ml}_j\over \partial J_{\nu}} &=&
{1\over \pi} \int\limits_{0}^{\pi} \biggl \{
{\partial \psi^{|m|}_j \over \partial R}{\partial R \over
 \partial J_{\nu}}\cos [l\theta _R+m(\theta _\phi -\phi)] \nonumber \\
&{}& \qquad -m\psi^{|m|}_j(R){\partial \over \partial J_\nu}
     \left(\theta_{\phi}-\phi\right)
     \sin [l\theta _R+m(\theta _\phi -\phi)] \biggr \}
 d \theta _R,~~\nu \equiv R,\phi. \label{eq::deriv-fourier-coeff}
\end{eqnarray}
Jalali \& Hunter (2005b) encountered these partial derivatives
in their second order perturbation theory devised for computing
the energy of eigenmodes. I adopt their technique for calculating
the quantities $\partial R/\partial J_{\nu}$ and
$\partial \left( \theta_{\phi}-\phi \right )/\partial J_{\nu}$.
The variables $R$, $(\theta _\phi -\phi)$, and $p_R$ are regarded
as functions of $(J_R,J_{\phi},\theta_R)$ because the action-angle
transformation
$(\textbf{\textit{x}},\textbf{\textit{p}})\rightarrow
(\textbf{\textit{J}},\Theta)$
is defined in the phase space of an axisymmetric state. From
$v_R=dR/dt=(\partial R/\partial \theta_R)\Omega_R$ one may write
\begin{equation}
\label{eq::dRdJ}
\frac{ d }{ d t}\left[{\partial R \over \partial J_\nu}\right]=
{\partial^2 R \over \partial \theta_R \partial J_\nu}\frac{d\theta_R}{dt}
=\Omega_R\frac{\partial}{\partial J_\nu}
\left({\partial R \over \partial \theta_R}\right)
=\Omega_R\frac{\partial}{\partial J_\nu}\left(\frac{v_R}{\Omega_R}\right)
={\partial v_R \over \partial J_\nu}-\frac{v_R}{\Omega_R}
{\partial \Omega_R \over \partial J_\nu}.
\end{equation}
Similarly, one obtains
\begin{eqnarray}
\frac{ d }{ d t}\left[{\partial \over \partial J_\nu}
 (\theta _\phi -\phi)\right]
&=& {\partial \Omega_{\phi} \over \partial J_\nu}
-\frac{\delta_{\nu,\phi}}{R^2}
+{2J_{\phi} \over R^3}{\partial R \over \partial J_\nu}
-\frac{1}{\Omega_R}\left[\Omega_{\phi}-\frac{J_{\phi}}{R^2}\right]
{\partial \Omega_R \over \partial J_\nu}, \label{eq::dthetadJ} \\
\frac{ d }{ d t}\left[{\partial v_R \over \partial J_\nu}\right]
&=& {2J_{\phi} \over R^3}\delta_{\nu,\phi}
-\left[{3J_{\phi}^2 \over R^4}+V_0^{\prime\prime}(R)\right]
{\partial R \over \partial J_\nu}
-\frac{1}{\Omega_R}\left[{J_{\phi}^2 \over R^3}-V_0^{\prime}(R)\right]
{\partial \Omega_R \over \partial J_\nu}. \label{eq::dvdJ}
\end{eqnarray}
The set of three equations (\ref{eq::dRdJ}) through (\ref{eq::dvdJ})
can be integrated along an orbit, and they
provide the additional values needed to evaluate
the partial derivatives (\ref{eq::deriv-fourier-coeff}).
Initial values are $\partial v_R/\partial J_\nu=
\partial(\theta_{\phi}-\phi)/\partial J_\nu=0$ at $\theta_R=t=0$
where $R=R_{\rm min}$ because $v_R=\theta_{\phi}-\phi=0$ for
all orbits. However the initial $R_{\rm min}$ values change with the
actions, and initial values for the derivatives of $R$ with respect
to the actions are
\begin{equation}
\left[ {\partial R \over \partial J_R}\right]_{R=R_{\rm min}}=
\frac{R^3_{\rm min}\Omega_R}
{R^3_{\rm min}V_0^{\prime}(R_{\rm min})-J^2_{\phi}}, ~~
\left[ {\partial R \over \partial J_{\phi}}\right]_{R=R_{\rm min}}=
\frac{R_{\rm min}(R^2_{\rm min}\Omega_{\phi}-J_{\phi})}
{R^3_{\rm min}V_0^{\prime}(R_{\rm min})-J^2_{\phi}}.
\end{equation}
They are obtained by differentiating the zeroth order energy equation.

\end{document}